\documentclass[aps,pra,groupedaddress, twocolumn,nofootinbib,superscriptaddress]{revtex4-1}
\newcommand{\be}{\begin{eqnarray}}
\newcommand{\ee}{\end{eqnarray}}
\newcommand{\beq}{\begin{equation}}
\newcommand{\eeq}{\end{equation}}

\usepackage{graphicx,epsfig}
\usepackage{amsmath}
\usepackage{amsfonts}
\usepackage{fancyhdr}
\usepackage{amsmath}
\usepackage{mathrsfs}
\usepackage[utf8]{inputenc} 
\usepackage[T1]{fontenc}
\usepackage{xcolor}

\newcommand{\Ket}[1]{\left| #1 \right>}\newcommand{\bra}[1]{\left< #1 \right|}

\usepackage{graphicx}

\begin{document}



\title{Static properties of two linearly coupled discrete circuits}

\author{Albert Escriv\`a}
\email{albert.escriva@fqa.ub.edu}
\affiliation{Departament de F\'{\i}sica Qu\`antica i Astrof\'{\i}sica, 
Facultat de F\'{\i}sica, Universitat de Barcelona, Mart\'{\i} i Franqu\`es 1, 08028 Barcelona, Spain}
\affiliation{Institut de Ci\`encies del Cosmos, Universitat de Barcelona,
08028 Barcelona, Spain}

\author{Andrea Richaud}
\affiliation{Scuola Internazionale Superiore di Studi Avanzati (SISSA), Via Bonomea 265, I-34136, Trieste, Italy}

\author{Bruno Juli\'a-D\'{\i}az}
\affiliation{Departament de F\'{\i}sica Qu\`antica i Astrof\'{\i}sica, 
Facultat de F\'{\i}sica, Universitat de Barcelona, Mart\'{\i} i Franqu\`es 1, 08028 Barcelona, Spain}
\affiliation{Institut de Ci\`encies del Cosmos, Universitat de Barcelona, 
08028 Barcelona, Spain}

\author{Montserrat Guilleumas}
\affiliation{Departament de F\'{\i}sica Qu\`antica i Astrof\'{\i}sica, 
Facultat de F\'{\i}sica, Universitat de Barcelona, Mart\'{\i} i Franqu\`es 1, 08028 Barcelona, Spain}
\affiliation{Institut de Ci\`encies del Cosmos, Universitat de Barcelona, 
08028 Barcelona, Spain}
%
\begin{abstract}
Bosonic two-ring ladders constitute an important class of atomtronic circuits, where coherent current flows not only can offer a new insight into many-body physics, but also can play the role of actual degrees of freedom, and hence allow for a viable implementation of cold-atom based devices and qubit systems. In this work, we exhaustively investigate the ground state properties and the low-lying energy spectrum of two linearly coupled Bose-Hubbard rings. We show that the competition among interactions, intra- and inter-ring hopping processes gives place to a rather rich physical scenario, where Mott-like states and (different kinds of) superfluid-like states emerge. The latter ones depend also on the (in)commensurate filling of the atoms. Our analysis, carried out within a simple analytical framework and by means of the exact numerical diagonalization of the system Hamiltonian, provides one with a rather complete characterization of the static properties of the two-ring ladder, including, but not limited to, coherence, fragmentation, correlations, and entanglement. We complement our investigation by studying how these indicators depend on the commensurability of the total number of bosons with respect to the total number of sites and show that the two stacked rings are always entangled for an odd number of atoms.  
%
%
%
%
%
%
\end{abstract}
%
\maketitle
\section{Introduction}

Ultracold quantum gases in closed geometries have become a basic building block of quantum
technologies. 
The versatility and high degree of experimental control over the interactions and the
geometry of ultracold atoms have made it possible to engineer 
atomic devices with different types of geometries and couplings, in particular,
rings and ladders.
This experimental progress has boosted the appearance of the emergent 
field of Atomtronics in quantum technologies~\cite{atomtronics,atomtronicexperimental1,REFEXTRA3,REFEXTRA6,Amico_Roadmap}.
The aim is to design
atomtronic circuits by coupling 
simple elements capable of producing complex applications with matter waves. 
For instance, quantum transport through different parts of the atomtronic circuit~\cite{REFEXTRA8},
the emulation of superconducting flux qubits used in quantum computing~\cite{Chiorescu},
or production of entangled states.
Entanglement is an important feature of quantum systems, 
that can be used for quantum information processing. 
Atomtronic devices may then be used to simulate intricate quantum 
systems~\cite{quantumsimulation}, to develop finer sensors and improve our 
metrological capabilities~\cite{atomtronics,REFEXTRA2,REFEXTRA7}.

The characterization of  ground state properties, correlations, entanglement and energy spectrum 
of ultracold bosons in a ring lattice with few sites
has been extensively investigated in the literature in different physical situations. For instance,
with contact interacting repulsive bosons ~\cite{Wu2006,Morales-Molina2012}, and
with attractive interactions~\cite{Buonsante2006}; 
the effects of dipolar interaction~\cite{Gallemi2013,Anna2013}, as well as
two distinct atomic species~\cite{Morales-Molina2012}. 
Moreover, the effects of a tunable tunneling in one  ring has been also addressed~\cite{gallemi1,gallemi2}.
Quantized vortices, which are characteristic flux states of superfluid systems in closed geometries,
 have been also explored in a ring lattice, see 
Refs.~\cite{Hallwood2006,Lee2006,Arwas2014,gallemi1,gallemi2,Paraoanu2003,Antonio2019}
and references therein.


In this paper we consider a primary integrated atomtronic circuit composed by two identical 
rings linearly coupled forming a two-ring ladder.
In particular, we investigate two identical 
trimers coupled by an 
inter-ring tunneling parameter, that can be different from the intra-ring tunneling 
between neighboring sites in the same ring. 
We identify the effects arising from the competition among interactions and the two tunnel couplings in order
to show that this is a flexible configuration that can be used
for quantum technologies. By properly tuning interactions and tunneling strengths, in fact, the system ground-state is shown to approach different notable configurations, which, in turn, constitute \textit{prototypical} atomtronic devices: from \textit{two stacked rings}, to \textit{two fully disconnected rings} or \textit{three disconnected double-wells}, as well as all the intermediate situations. 

The two-stacked-ring geometry is at the base of experimentally-available implementations of a matter-wave based qubit system \cite{amico1,amico3}. In these platforms, which combine the long coherence lifetime of neutral cold atoms systems with the robustness of topologically protected solid state Josephson flux qubits \cite{Mooij_qubit}, it has been shown that the associated imaginary-time effective action provides a two-level-system dynamics for the phase slip across the two rings \cite{amico1}. From the point of view of many-body physics, in the presence of a synthetic magnetic field, weakly- and strongly-interacting bosons on two-leg-ladder geometries are well known to disclose intriguing phase diagrams and complex magnetic-like phenomena, where circulating currents can exhibit characteristic Meissner-like and vortex-like patterns 
\cite{PhysRevB.92.060506,Orignac_Meissner,Orignac_Incommensurate,Orignac_vortex,Citro_QPT,felipe,amico2,REFEXTRA4,Victorin2018}, reminiscent of the field dependence of currents in type-II superconductors \cite{piraud}. In addition, it is worth mentioning that a rich dynamical scenario emerges if one focuses on the possible transfer of persistent currents and vortices between linearly coupled rings \cite{Oliinyk_Symmetry,Oliinyk_JPB,Oliinyk_Nonlinear,andrea,lesanovsky,Nicolau,albert_dynamic,gallemi3}.

When the inter-ring tunneling is negligible with respect to the intra-ring 
hopping, the two rings are effectively decoupled. The resulting single-ring system has been subject of extensive study \cite{Amico_Roadmap}, as it can support the realization of a supercurrent-based qubit \cite{amico3,REFEXTRA1}.
In the opposite limit, that means when it is the intra-ring tunneling to be negligible with respect to the inter-ring tunneling, one remains with many decoupled double-well systems. The latter constitute the fundamental building block of several atomtronic devices \cite{Obertaler,Schumm,Spagnolli,Valtolina_Josephson, Burchianti_Josephson}, as it is well known that a Bose-Einstein condensate of neutral atoms in a double-well potential represents the matter-wave counterpart of a Josephson junction of coupled superconductors \cite{Amico_Roadmap}.

%
%


Depending on the relative value of its parameters, the model we investigate thus interpolates among different prototypical atomtronic systems and can hence capture not only standard ``asymptotic limits", but also interesting cross-over regimes, where the competition among different couplings is at its most crucial. The aim of this paper is hence to provide an exhaustive characterization of the ground state properties of two coupled rings, in a wide range of tunneling strengths and interactions. We use exact diagonalization techniques to characterize the ground state properties of the two-ring bosonic ladder as a function of the tunneling parameters and onsite interaction strength for different number of atoms. For instance, we investigate fragmentation, entanglement and quantum correlations properties of this elemental integrated circuit that can help to guide future applications in atomtronics technologies. We pay special attention to the commensurability or incommensurability of the number of atoms with respect to the total number of sites of the system.  We discuss the main differences, and we show that they could be used to indirectly determine whether the number of trapped bosons is even or odd. For instance, when the number of atoms is odd, the two rings remain always entangled for any range of onsite interactions.  We also provide analytical expressions in the limit of large interactions in the case of non commensurate number 
of atoms.
These characteristic features of a quantum system are particularly relevant as recent experiments are already able  to measure entanglement properties~\cite{entropy-experimental} and quantum  correlations in these setups~\cite{correlation-experimental,REFEXTRA9}.

Our paper is organized as follows: In Sect.~\ref{sec2} we introduce the model 
Bose-Hubbard (BH) Hamiltonian as well as the parameters of the physical system. 
In Sect.~\ref{sec3} we present the analytical solutions of our model 
for the single-particle case.
%
%
We discuss also the properties of the eigenstates 
of the non-interacting system and the low-lying energy spectrum obtained numerically.
In Sect.~\ref{sec4} we consider $N$ bosons loaded in the two-ring lattice 
and, by numerical diagonalization of the Hamiltonian, 
we investigate the static properties of the ground state of the two-coupled rings
for different values of $N$, as well as the effects of the non commensurability of the 
number of atoms with respect to the number of sites, 
such as fragmentation, correlations between different sites, Schmidt gap and the entanglement entropy. 
Finally, in Sect.~\ref{sec6} we provide the main conclusions of our work.

\section{Theoretical model}
\label{sec2}

\begin{figure}[t]
\includegraphics[width=.6\columnwidth]{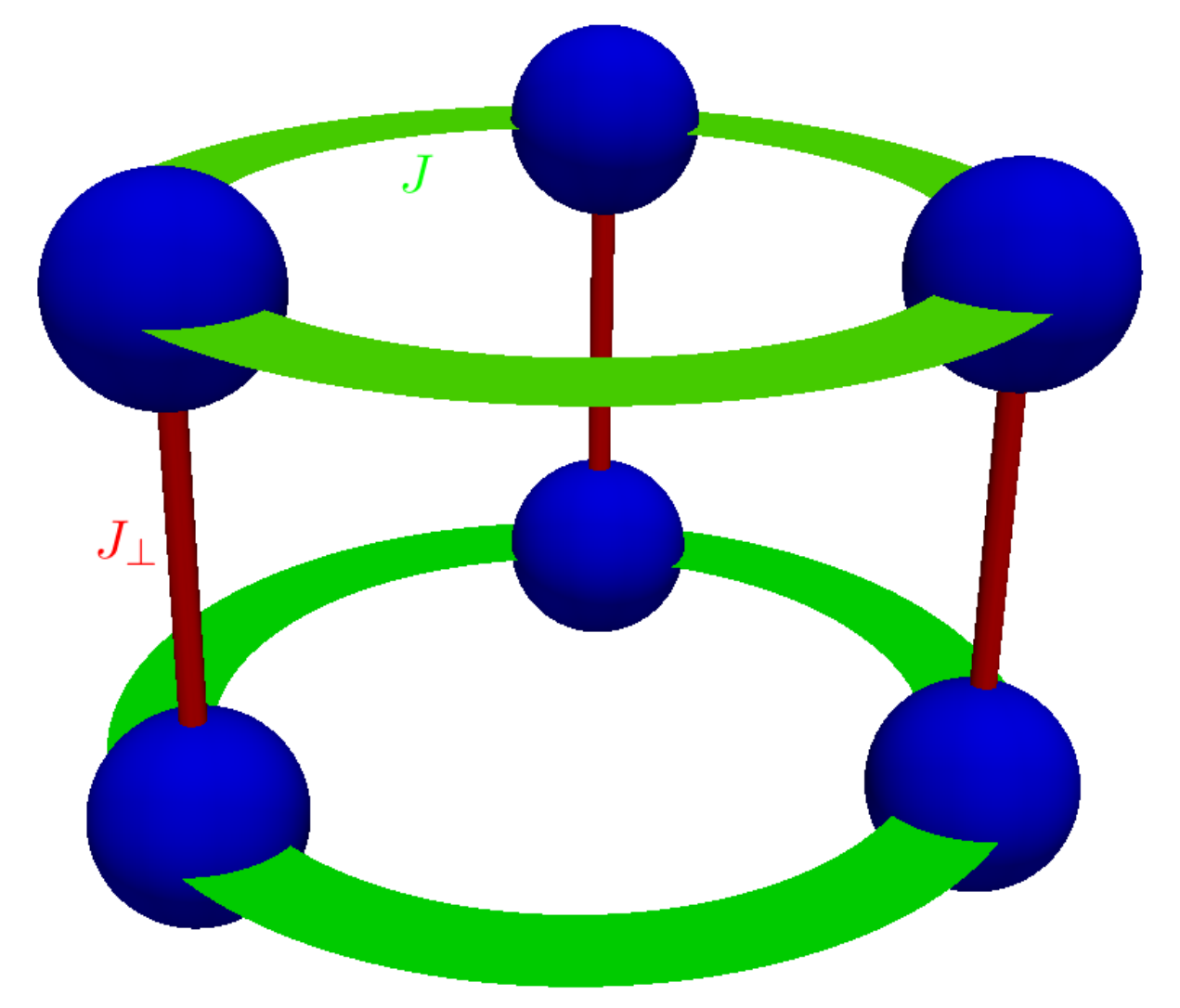} 
\caption{Schematic representation of the minimal system with two stacked trimers. 
The intra-ring tunneling between the sites of the same ring is given by $J$ and 
the inter-ring tunneling  between sites of different rings is given by $J_{\perp}$.}
\label{Fig-esquema}
\end{figure}

We consider $N$ bosons loaded in two stacked Bose-Hubbard rings, 
with the same number of sites $M$ in each ring and tunnel coupled via the rungs, see Fig.~\ref{Fig-esquema} for a schematic representation. The system is described by the following Hamiltonian,
\be
\label{hamiltonian}
 \mathcal{\hat{H}} &&=  
- J \sum_{j=\uparrow,\downarrow}
\sum_{i=1}^{M}  \,(\hat{a}^{\dagger}_{i,j} \, \hat{a}_{i+1,j}+\hat{a}^{\dagger}_{i+1,j} \,\hat{a}_{i,j} )\\
&&- J_{\perp} 
\sum_{i=1}^{M} \,(\hat{a}^{\dagger}_{i,\uparrow}\,\hat{a}_{i,\downarrow} 
+ \hat{a}^{\dagger}_{i,\downarrow}\,\hat{a}_{i,\uparrow})
+\frac{U}{2}\,\sum_{j=\uparrow,\downarrow}\sum_{i=1}^{M}\hat{n}_{i,j}(\hat{n}_{i,j}-1)\nonumber  \,,
\ee
where $J$ and $J_{\perp}$ are the tunneling parameters between neighboring sites 
in the same ring, and between the two rings, respectively. The bosonic creation 
(annihilation) operators $\hat{a}^{\dagger}_{i,j}$ ($\hat{a}_{i,j}$) for the site $i$ 
in the ring $j$ fulfil the canonical commutation relations, 
$[\hat{a}_{i,j},\hat{a}^{\dagger}_{k,l}]=\delta_{ik}\delta_{jl}$. We introduce also 
the usual particle number operators $\hat{n}_{i,j}=\hat{a}^{\dagger}_{i,j} \hat{a}_{i,j}$ 
of the $i$th site of the ring $j$. $U$ sets the strength of the atom-atom 
interaction which is assumed to be repulsive ($U>0$).

This coupled system can be interpreted as a two-leg Bose-Hubbard ladder with 
periodic boundary conditions forming two connected rings with $M$ sites each 
one. In the ladder geometry,  $J$ and $J_{\perp}$ correspond to the hopping along 
the rings and legs of the ladder. The competition between the two tunneling parameters determines the static 
properties of the two connected discrete rings. In particular, there are two 
limiting cases: when $J/ J _{\perp}\rightarrow 0$ the system behaves as $M$ 
independent bosonic Josephson junctions, with two sites coupled by 
$J_\perp$~\cite{junction}, whereas when $J/ J _{\perp}\rightarrow \infty$ the two 
rings become fully decoupled. Many-body properties of a single lattice ring with finite 
number of sites loaded with contact interacting bosons,
have been studied in Refs.~\cite{Wu2006,Buonsante2006}. A general case of one ring
with  a tunable tunneling, has been previously studied for $M=3$~\cite{gallemi1} 
and for an arbitrary (but small) 
number of sites~\cite{gallemi2}. In this work we consider mainly the minimal 
coupled atromtronic circuits; two stacked trimers.

\section{Non-interacting limit}
\label{sec3}

In this section we revisit the non-interacting limit of Hamiltonian (\ref{hamiltonian}), which admits analytical solutions. We will focus on the case $M=3$, which constitutes the minimal two-ring configuration.

The non-interacting solutions of two linearly coupled trimers can be constructed as,
%
%
\begin{equation}
\Ket{\Psi_{q}^{\pm}}=\frac{1}{\sqrt{2 M}} \, 
\sum_{l=1}^{M} e^{i \frac{2 \pi q \,l}{M}} \,(\hat{a}^{\dagger}_{l,\uparrow} 
\pm \hat{a}^{\dagger}_{l,\downarrow} ) \Ket{\rm vac} \, ,
\label{ev-2rings}
\end{equation}
where $\Ket{\rm vac}$ stands for the vacuum, $\hat{a}^{\dagger}_{l}$ is  the creation operator of one atom in the $l$ site, and $q=0,1,...,M-1$ labels the vortex wave function with quantization $2 \pi q$.

This solution was also found in Ref.~\cite{cosafalsa} but in the continuous limit 
for an infinite number of sites by diagonalizing the Hamiltonian in momentum space by means of a Bogoliubov transformation. It is interesting to note that for the single-particle case, 
the eigenstates of two-coupled rings are independent of the 
tunneling parameters $J$ and $J_\perp$, and they involve only 
the same kind of flux state in both rings simultaneously. This 
combination can be either symmetric or antisymmetric. 
This means that a state formed by a combination of a vortex in the top ring and 
an antivortex in the bottom ring is not an eigenstate of the non-interacting two-coupled ring 
system.

In Ref.~\cite{albert_dynamic} we have studied linear combinations of non-interacting 
degenerate states, when $J_\perp=0$, that lead to
single-particle vortices and fractional vortices. These states correspond to  
population-imbalanced vortices with current localized in one ring, and to 
population-balanced fractional vortices, respectively.
They are stationary in the noninteracting limit of decoupled rings.

\begin{figure}[t]
\includegraphics[width=1\linewidth]{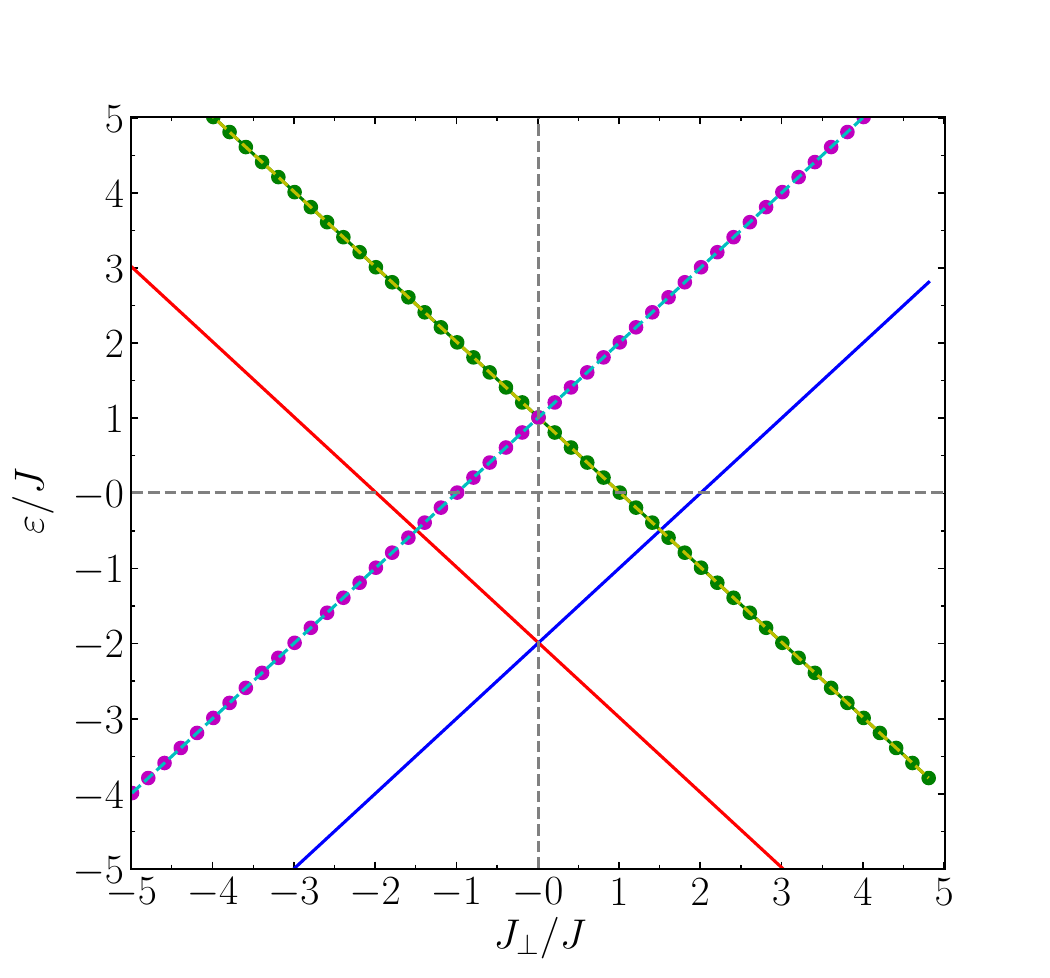} 
\caption{Energy spectrum of two coupled trimers in the single-particle 
case as a function of $J_{\perp}/J$. Red line corresponds to $\varepsilon_{0}^+$, 
blue line $\varepsilon_{0}^-$, green-dotted line  $\varepsilon_{\rm v}^+$, yellow-dashed 
line $\varepsilon_{\rm av}^+$, magenta-dotted line $\varepsilon_{\rm v}^-$ and 
blue-dashed line  $\varepsilon_{\rm av}^-$.}
\label{eigenvalues}
\end{figure}

In Fig.~\ref{eigenvalues} we show the single-particle energy spectrum for 
two stacked trimers as a function of the ratio between the inter-ring and 
intra-ring couplings $J_\perp/J$. The ground state is non degenerate for all 
values of the tunneling except for $J_\perp=0$, when there is no coupling 
between the two rings. In this case the two rings are independent and 
therefore the symmetric and antisymmetric combinations have the same energy.
When $J_{\perp}/J>0$ ($J_{\perp}/J<0$) the ground state is the symmetric 
(antisymmetric) combination of the single ring ground state solutions with 
energies $\varepsilon_{0}^+= -J(J_{\perp}/J+2)$ and $\varepsilon_{0}^-= J(J_{\perp}/J-2)$, 
respectively. The symmetric and antisymmetric excited states present a 
double degeneracy corresponding to vortex-vortex and antivortex-antivortex 
combinations: $\varepsilon_{\rm v}^+ =\varepsilon_{\rm av}^+=J(1-J_{\perp}/J)$ 
and $\varepsilon_{\rm v}^-=\varepsilon_{\rm av}^-=J(1+J_{\perp}/J)$, where we have 
used the notation $q=0 \,({\rm gs}),1 \,({\rm v}),-1 \,({\rm av})$. The energy 
spectrum is symmetric with respect to the change of sign of $J_{\perp}/J$. %

In order to study the static properties of our system for different values of 
the interaction and tunneling parameters, we calculate the ground state and 
the lower part of the energy spectrum by numerical diagonalization of the 
Hamiltonian~\cite{exact-diagonalization,Bruno1} for different numbers of atoms.

\begin{figure}[t]
\includegraphics[width=1\linewidth]{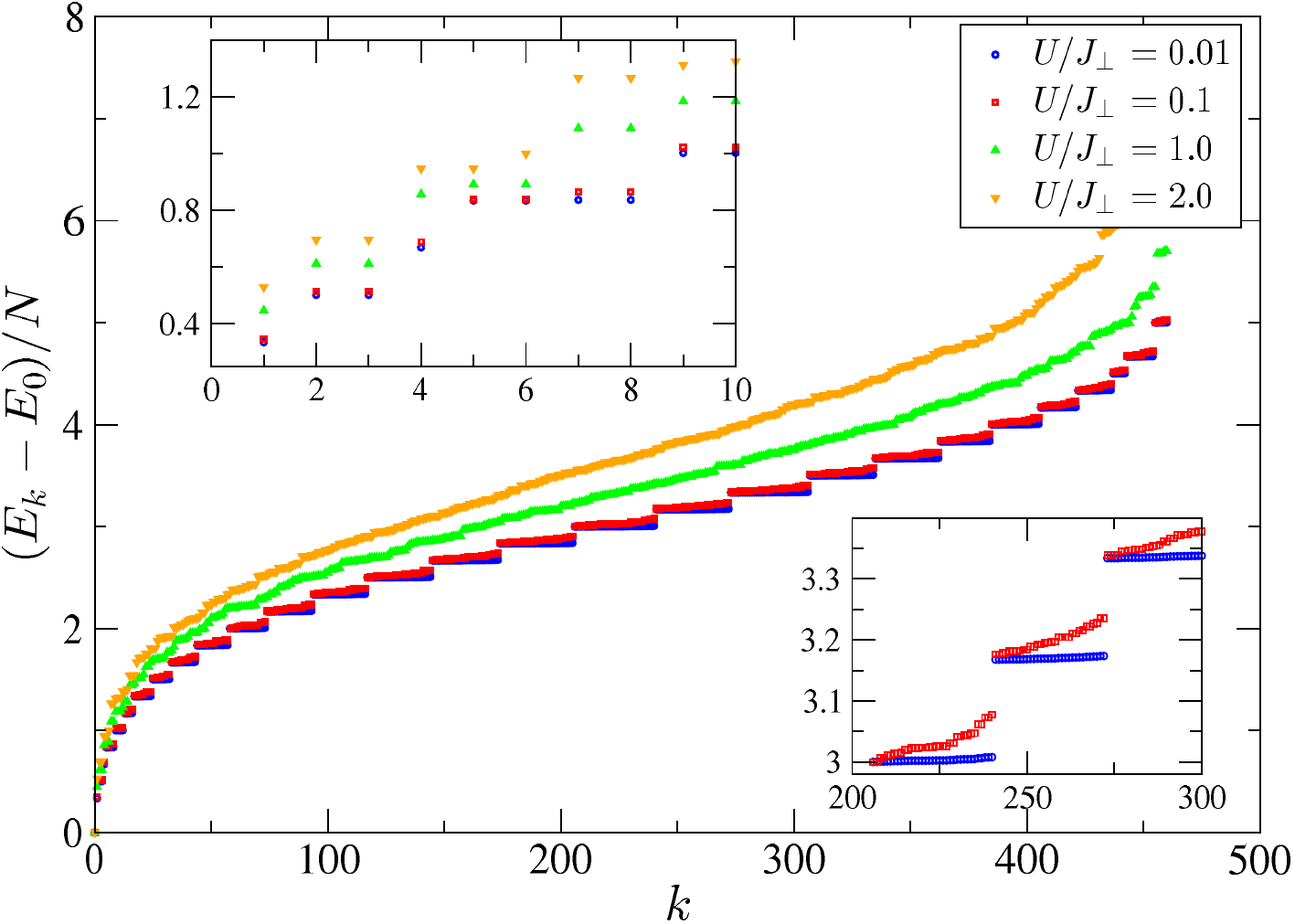} 
\caption{Excitation energies per particle, $(E_{k}-E_{0})/N$ of the system with $N=6$, $M=3$ and $J/J_{\perp}=1.0$, for different values of $U/J_\perp$. }
\label{spectrum_U}
\end{figure}

In Fig.~\ref{spectrum_U} we plot the energy spectrum, for 
different values of  the interaction $U/J_{\perp}$, where $k$ is the spectral index.
We observe the existence of energy bands for small values of the interaction, which breakdown when the 
interaction rate is increased. 
As explained in Ref.~\cite{gallemi2} the band structure can be understood by means of the number of atoms
and the degeneracy of the flow basis.
%
%

\section{Ground state properties of the interacting system}
\label{sec4}

The system formed by $N$ interacting bosons loaded in two coupled atomtronic circuits is described by the 
Hamiltonian (\ref{hamiltonian}) and has three physical parameters: the interatomic 
interaction $U$, and the tunneling strengths $J$ and $J_\perp$. In modern experimental setups,
these parameters can be tuned in a very controlled way \cite{amico3,amico4}, although their specific variability ranges strictly depend on the chemical element which is Bose-condensed and on the optical apparatus which is employed. In Ref.~\cite{chiral_antonio}, for example, the experimental realization of a bosonic two-leg ladder in the \textit{weakly} interacting regime, meaning that $U/J_\perp \approx 0$, was reported. In this remarkable experiment, measurements were performed in the range $J_\perp/J \in [0,\,3.5]$, i.e. from a regime where the two rings are decoupled to a regime where the inter-ring coupling is $3.5$ times larger than the intra-ring coupling. In Ref.~\cite{piraud}, the complementary regime was explored, i.e. the \textit{strongly} interacting one (which was claimed to be within the reach of available experimental setups). 
Numerical simulations were performed in a rather extended range of model parameters, i.e. $U/J\in[0,\,+\infty)$ and $J_\perp/J \in [0,\,8]$.

In this section, we characterize the 
ground state of the two-connected trimers by means of coherence and fragmentation 
properties, correlations between different sites, as well as the entanglement between the two rings.
We will discuss also the effect of incommensurate filling.

\subsection{Limiting situations}

Let us recall the analytical solutions of the ground state in two limiting cases.
%
When the tunneling terms dominate ($U/J \rightarrow 0$ and $U/J_\perp \rightarrow 0$),
this leads to a complete delocalization of each atom over all the sites. 
The non-interacting limit in the homogeneous case corresponds to the 
superfluid phase (Bose-Einstein condensate, BEC). In our discrete system, 
the coherent ground state reads,
\begin{equation}
\Ket{\Psi_{\rm BEC}}= 
\frac{1}{\sqrt{N!}}\left(\frac{1}{\sqrt{2M}}\sum_{j=\uparrow,\downarrow} 
\sum_{i=1}^{M}\hat{a}^{\dagger}_{i,j} \right)^{N} 
\Ket{\rm vac} \,.
\label{wf-BEC}
\end{equation}
When the repulsive interaction dominates ($J/U \rightarrow 0$ and $J_\perp/U \rightarrow 0$, with $U>0$)
the tunneling energies become small and the atoms tend to localize.
Therefore, the system prefers to reduce the number of pairs in 
each site to minimize the energy. The ground state has equipopulation with, 
on average, $\nu=N/(2M)$ atoms on each site, when $\nu \in \mathbb{Z}$. 
The strong interacting limit of one ring with incommensurate filling 
(non integer filling factor) has been studied in Ref.~\cite{Wu2006}.
Here we start the analytical study to commensurate number of atoms, but we also
discuss the ground state properties of two coupled rings when one atom is added (or subtracted)
from a commensurate case.
The localized ground state for a commensurate system (Mott insulator phase, MI, in the homogenous system) corresponds to 
one state of the Fock basis:
\begin{equation}
\Ket{\Psi_{{\rm MI}, \nu}} = \prod_{j=\uparrow,\downarrow}  
\prod_{i=1}^{M} \frac{(\hat{a}^{\dagger}_{i,j})^{\nu}}{\sqrt{\nu!}} \Ket{\rm vac},
\label{wf-MI}
\end{equation}
assuming that the number of atoms $N$ is commensurate with the number of sites $2M$,
$\nu=N/(2M) \in \mathbb{Z}$.
When one atom is added (or subtracted) the system has $N\pm 1$ atoms with filling factor
$\nu^{\pm} = (N \pm 1)/(2M)=\nu \pm 1/(2M)$.
The ground state of the resulting system is, respectively:
\begin{eqnarray}
\label{wf-MIextra}
\Ket{\Psi_{{\rm MI}, \nu^+}} &=& \frac{1}{\sqrt{2M}}\frac{1}{\sqrt{\nu+1}}\sum_{j=\uparrow,\downarrow} 
\sum_{i=1}^{M} {\hat{a}}^{\dagger}_{i,j} 
\Ket{\Psi_{{\rm MI}, \nu}} \,, \nonumber \\
\Ket{\Psi_{{\rm MI}, \nu^-}} &=& \frac{1}{\sqrt{2M}}\frac{1}{\sqrt{\nu}}\sum_{j=\uparrow,\downarrow}   
\sum_{i=1}^{M} {\hat{a}_{i,j}} 
\Ket{\Psi_{{\rm MI}, \nu}} \,.
\end{eqnarray}
These two states can be obtained by distributing an extra particle (or vacancy) 
between the different sites of the commensurate state (\ref{wf-MI}).

\subsection{Coherence and fragmentation}
\label{ss:fc}

We investigate the coherence of the system by analyzing the 
condensed fraction and fragmentation properties which are defined 
by the eigenvalues of the one-body density matrix~\cite{condensed_fractions}. 
For a many-body state $\Ket{\Psi}$ describing $N$ bosons in $2M$ 
sites (two rings with $M$ sites), the one-body density matrix, $\hat{\rho}$, 
is a $2M \times 2M$ matrix, whose elements read
\begin{equation}
\rho_{(i,j),(k,l)}=\frac{1}{N}\bra{\Psi} {\hat a}^{\dagger}_{i,j} \, {\hat a}_{k,l} \Ket{\Psi},
\label{one-body}
\end{equation}
where $i,k =1,...,M$ and $j,l=\uparrow,\downarrow$. Since $\Ket{\Psi}$ is 
normalized to one, this implies that ${\rm Tr} (\hat \rho) =1$. The eigenvalues of 
the one-body density matrix are the relative occupation numbers, $p_{i}=N_{i}/N$, 
of the corresponding eigenvectors (single-particle states or natural orbitals). 
They are normalized to $1$,  meaning that $p_1 + p_2 + ..+ p_{2M}=1$, and they are conventionally labelled in such a way that $p_1 \ge p_2 \ge ..\ge p_{2M} \ge 0$
\cite{condensed_fractions}.
In the case of a singly condensed system there is only one large eigenvalue 
$p_{1} \approx 1$ and the others are very small $p_{i} \sim {\cal O}(1/N)$ ($i \neq 1$). 
This means that there is a macroscopic occupation of the corresponding  single-particle state $ \sim {\cal O}(N) $ and the system is condensed (it is worth 
stressing that the macroscopic occupation of a natural orbital is a typical feature of Bose-Einstein condensation and it is what supports the possibility of introducing a "macroscopic wave function" for the system \cite{condensed_fractions}.
Whereas when there is more than one large eigenvalue, the system is fragmented \cite{condensed_fractions}.
The ground state with largest fragmentation corresponds to all eigenvalues $p_i = 1/(2M)$
in the Mott insulator limit.

We remark that \textit{coherence} is a crucial ingredient in the design of atomtronic circuits \cite{Amico_Roadmap,Pepino}, while \textit{fragmentation} tends to hinder their operation. When the two-ring ladder is used as a superfluid qubit system \cite{amico3}, condensate's fragmentation tends to destroy the topologically protected quantum state (which realizes the qubit itself) and should therefore be minimized. In addition, fragmentation, which is inevitably present in interacting systems, can lead to the suppression of the coherent transfer
of vorticity between the rings \cite{albert_dynamic}.  

\begin{figure}[t]
\vspace*{-0.4cm}
\includegraphics[width=1\linewidth]{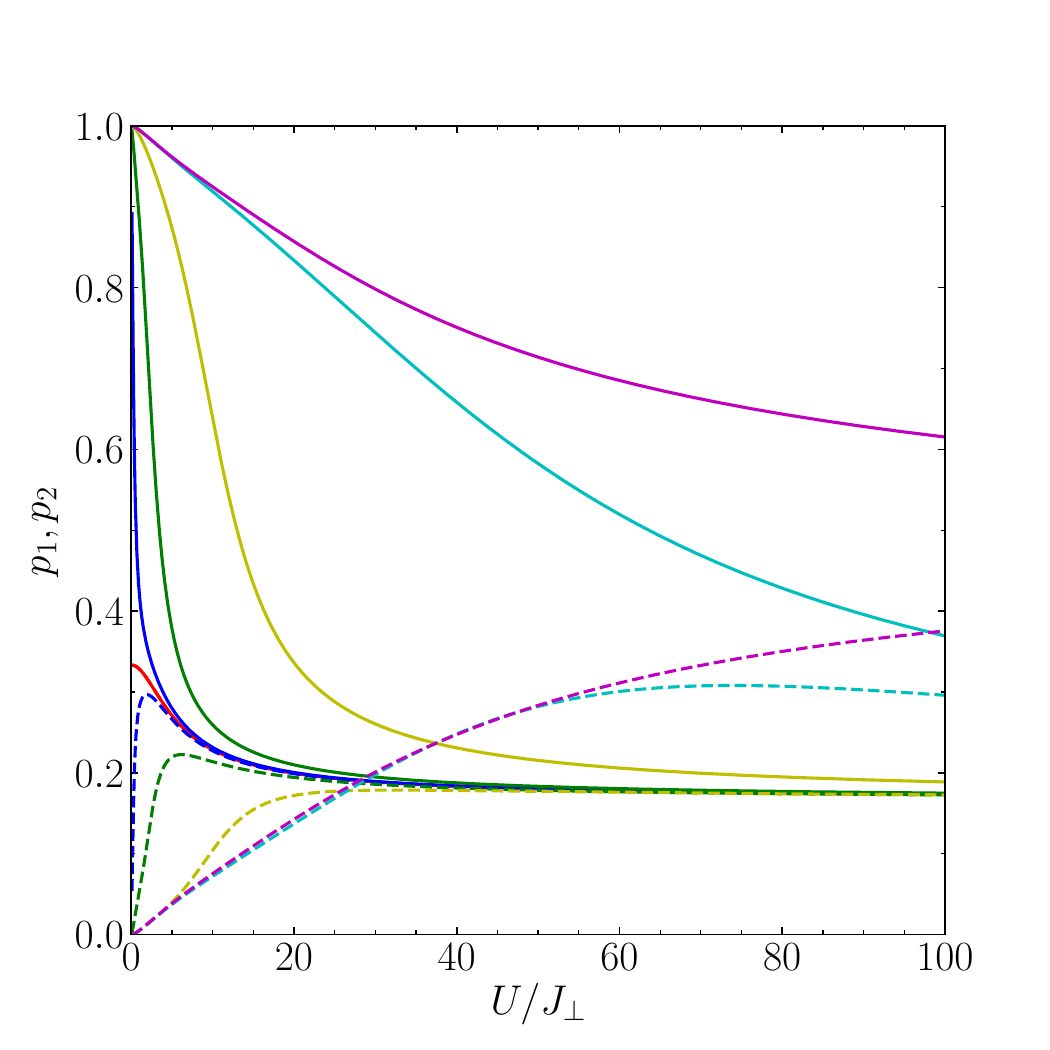} 
\caption{The two largest eigenvalues of the one-body density matrix, 
$p_{1}$ (solid) and $p_{2}$ (dashed), as a function of $U/J_{\perp}$ for 
different values of $J/J_{\perp}$. In all cases: $N=6$ and 
$M=3$. Color legend (from bottom to top): red ($J/J_{\perp}=0$, $p_{1}$ and $p_{2}$ are overlapped), blue 
($J/J_{\perp}=10^{-2}$), green ($J/J_{\perp}=10^{-1}$), yellow ($J/J_{\perp}=1$), 
cyan ($J/J_{\perp}=10^{1}$) and magenta ($J/J_{\perp}=10^{2}$).}
\label{eigenvalues_density_matrix}
\end{figure}
In Fig.~\ref{eigenvalues_density_matrix} we show the two largest relative 
occupation numbers, $p_{1}$ and $p_{2}$, of two coupled trimers as a function 
of $U/J_{\perp}$, with $N=6$ and for different values of $J/J_\perp$. 
For weak inter-particle interactions ($U \ll J_\perp$) 
there is a non-zero single particle state, $p_1 \simeq 1$, for all values of $J/J_\perp$ except when 
$J/J_{\perp}=0$. In the latter case, the sites in the same ring are fully 
decoupled and the system reduces to three independent double-wells with 
tunneling rate $J_{\perp}$. Thus, a three-fragmented state with $p_{1}=p_{2}=p_{3}=1/3$ 
is obtained when $J/J_\perp=0$ and $U/J_\perp \lesssim 1$. 
In this case the total wave function is not given by Eq.~(\ref{wf-BEC}), 
but by a product state of three independent bosonic Josephson junctions 
with $N/M$ particles in each junction, when $N/M \in \mathbb{Z}$.

As the interaction increases and the ratio $U/J_\perp$ becomes large, the atoms 
start to localize and the system tends to a Mott-insulator-like phase in the 
asymptotic limit where the ground state is fully fragmented and $p_{i}=1/(2M) = 1/6$.
However, due to the particular geometry of our system with two competing 
tunneling rates $J$ and $J_\perp$, there are two different regimes, $J>J_{\perp}$ 
and $J_{\perp}<J$, which clearly appear in the asymptotic values of 
Fig.~\ref{eigenvalues_density_matrix}. 
When $J/J_\perp \leq 1$ the ground state reaches a fragmented state for smaller 
values of $U/J_\perp$ than when $J/J_{\perp} > 1$. 
For a fixed value of $U/J_{\perp}$, when the intra-ring tunneling is dominant 
in front of the inter-ring coupling ($J>J_{\perp}$), the system is less fragmented 
than for $J/J_{\perp}<1$. This follows from the condition $\sum_{i}^{2M} p_{i} =1$. 
Figure~\ref{eigenvalues_density_matrix} shows that $p_1+p_2 \simeq 1$ when 
$J>J_{\perp}$ , whereas when $J<J_{\perp}$ all the eigenvalues of the one-body 
density matrix $p_i \neq 0$ which corresponds to a larger fragmentation of the 
ground state.
The system we are considering has $2M$ intra-ring couplings $J$
(that is, $2M$ pairs of sites connected by $J$)
and $M$ inter-ring couplings $J_{\perp}$
($M$ pairs of sites connected by $J_\perp$). Thus, when $J>J_{\perp}$ the tunneling effects 
are larger than when $J<J_{\perp}$, and the ground state remains delocalized 
for larger values of $U/J_\perp$. The system needs larger interactions to balance the
delocalization promoted by the tunneling. A similar reasoning holds also if one increases the (commensurate) number of atoms $N$. As is well known, in fact, higher (integer) fillings correspond to smaller Mott lobes and, therefore, to systems which undergo Mott localization for larger values of the interaction $U$ \cite{Mott_Superfluid_critical}. Hence, upon increasing the (integer) filling $\nu$, the behaviour illustrated in Fig. \ref{eigenvalues_density_matrix} is qualitatively unchanged, although the asymptotic value $p_1=p_2=1/(2M)$ is reached for \textit{larger} values of $U$.

A complementary information is provided by its corresponding von Neumann entropy, which 
measures the condensation of the system. It is defined from the eigenvalues 
of the one-body density matrix as
\begin{equation}
 S_{\rm VN}=-\sum_{i=1}^{2M} p_{i} \ln p_{i}  \,.
 \label{eq-S_VN}
\end{equation}
It is a bounded quantity: the minimum value is $S_{\rm VN}=~0$ and corresponds 
to a fully condensed ground state ($p_1= ~1$), whereas the maximum value
of $S_{\rm VN}$ is $\ln(2M)$ and occurs for a completely fragmented state with 
$p_i=1/(2M)$, for all $i$. When $S_{\rm VN} = \ln s$ (with $s \in \mathbb{Z}^+$) the 
system is fragmented in $s$ states.
\begin{figure}[t]
\includegraphics[width=1\linewidth]{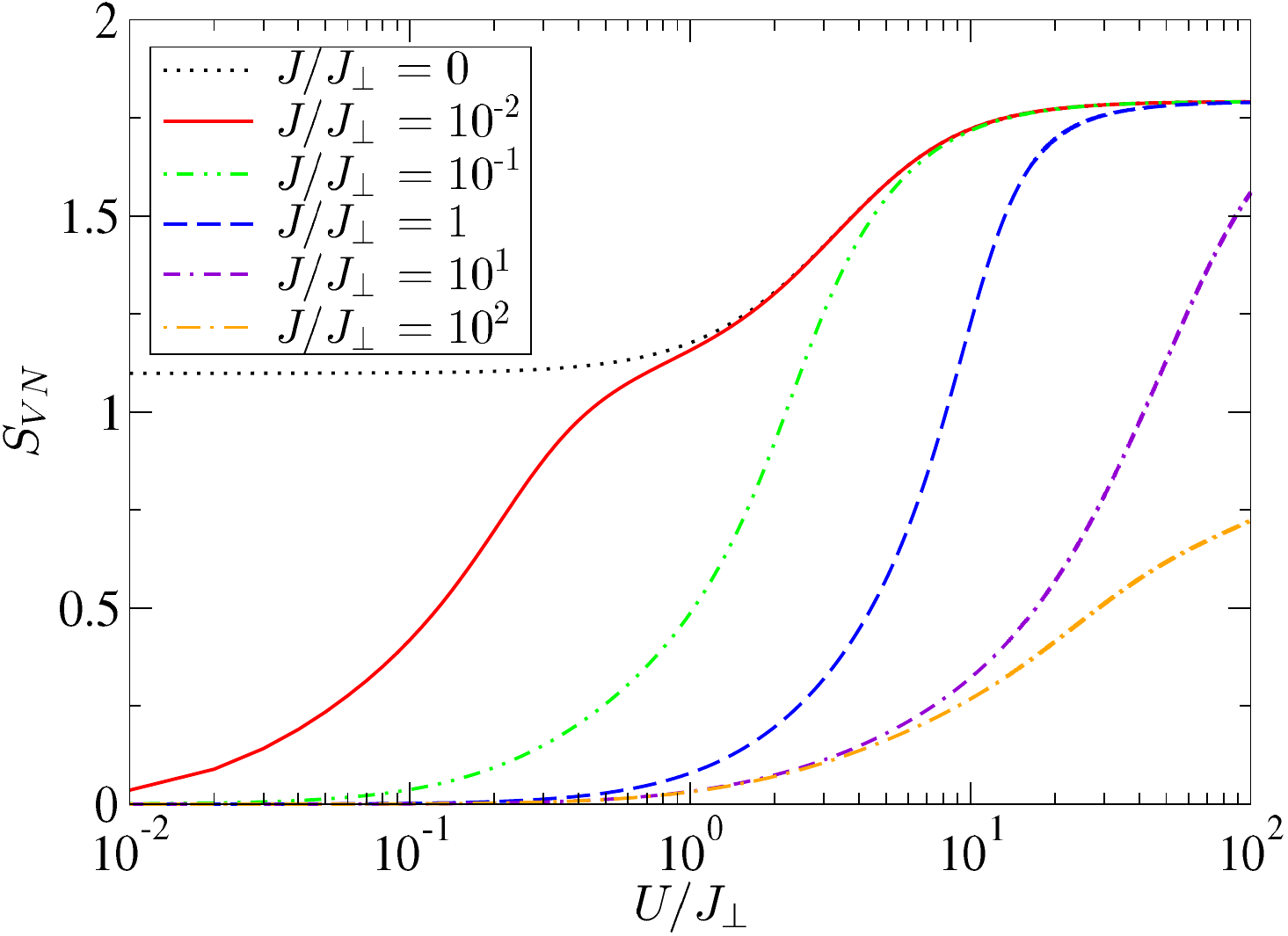} 
\caption{Von Neumann entropy of the one body density matrix as a function of $U/J_{\perp}$ for different 
values of $J/J_{\perp}$. In all cases, $N=6$ and $M=3$.}
\label{VNentropy}
\end{figure}
Figure~\ref{VNentropy} presents $S_{\rm VN}$ for the two stacked
trimers with $N=6$ atoms as a function  of $U/J_{\perp}$, for the same cases 
$J/J_\perp$ as in Fig.~\ref{eigenvalues_density_matrix}. 
One can see that for a fixed value of $U/J_\perp$, the entropy decreases as 
$J/J_\perp$ increases; that is, the ground state is less fragmented when 
$J>J_{\perp}$ than when $J<J_{\perp}$, as we have already obtained in 
Fig.~\ref{eigenvalues_density_matrix}. In the limiting case $J=0$ and when 
$U/J_{\perp}>1$ the entropy is given by $S_{\rm VN}=\ln M= \ln 3$, since the 
system behaves as $M=3$ bosonic Josephson junctions. 

\begin{figure}[t]
\includegraphics[width=1\linewidth]{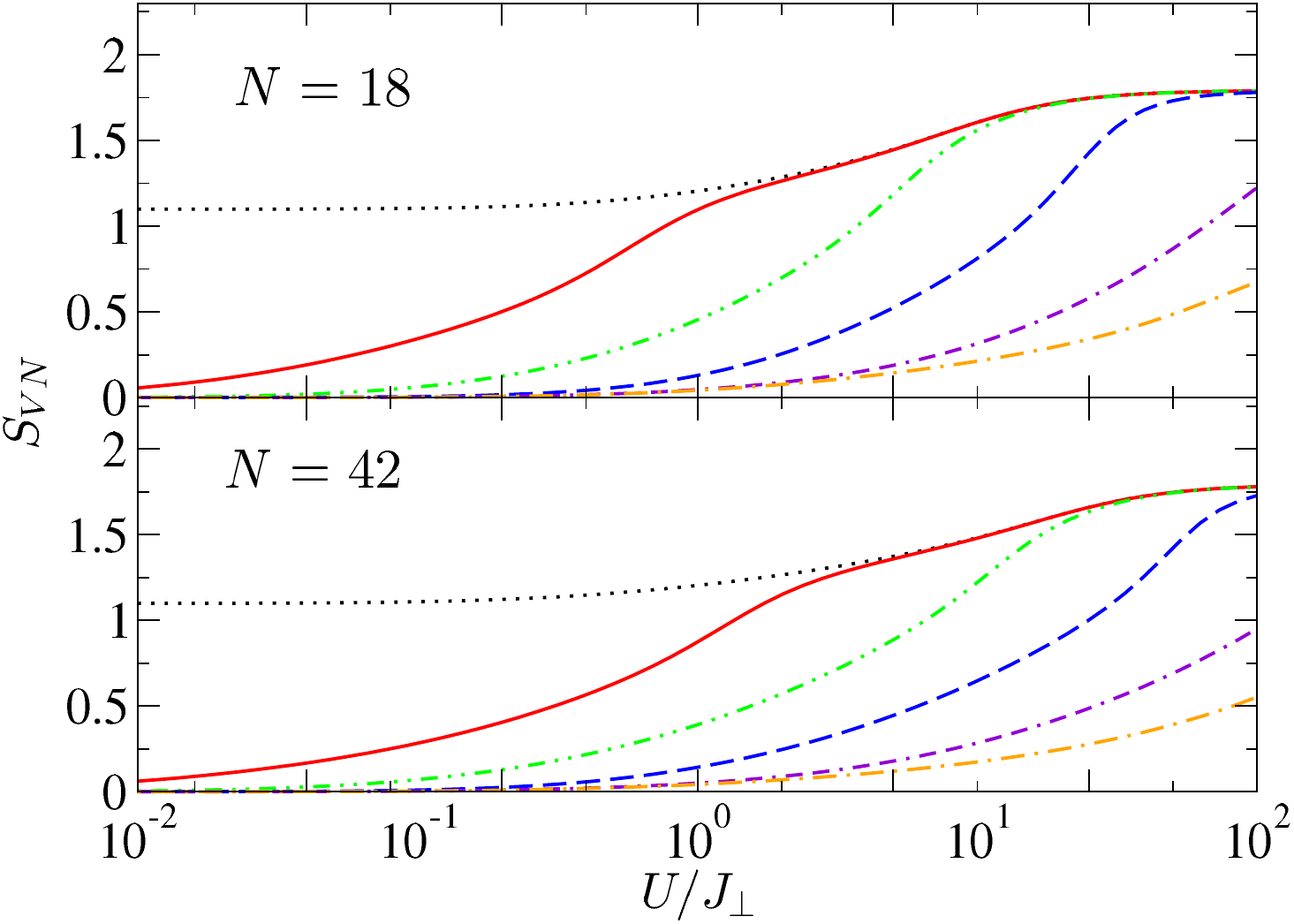} 
\caption{Von Neumann entropy of the one body density matrix as a function of $U/J_{\perp}$ for different 
values of $J/J_{\perp}$ and particles $N$: top panel ($N=18$),
bottom panel ($N=42$). In all cases $M=3$.}
\label{VNentropy_extra}
\end{figure}
When $J/J_\perp \geq 0.1$ and for small values of $U/J_\perp$, the ground state 
corresponds to a coherent state
and $S_{\rm VN} \simeq 0$. From Fig.~\ref{VNentropy} 
one can see that this regime in terms of $U/J_\perp$ broadens 
when $J > J_\perp$.
For large interactions the system tends to the hard-core boson limit with 
a $2M$ fragmented state, whose entropy is $S_{\rm VN}=\ln(2M) =\ln 6$. This 
regime is achieved for smaller values of $U/J_\perp$ when $J<J_\perp$. The 
localization of the particles in different sites  requires larger interaction 
strengths, $U/J_\perp > 10^2$ when $J/J_\perp >1$, as we have commented in 
the previous figure. 

\begin{figure*}[t]
\includegraphics[width=0.98\linewidth]{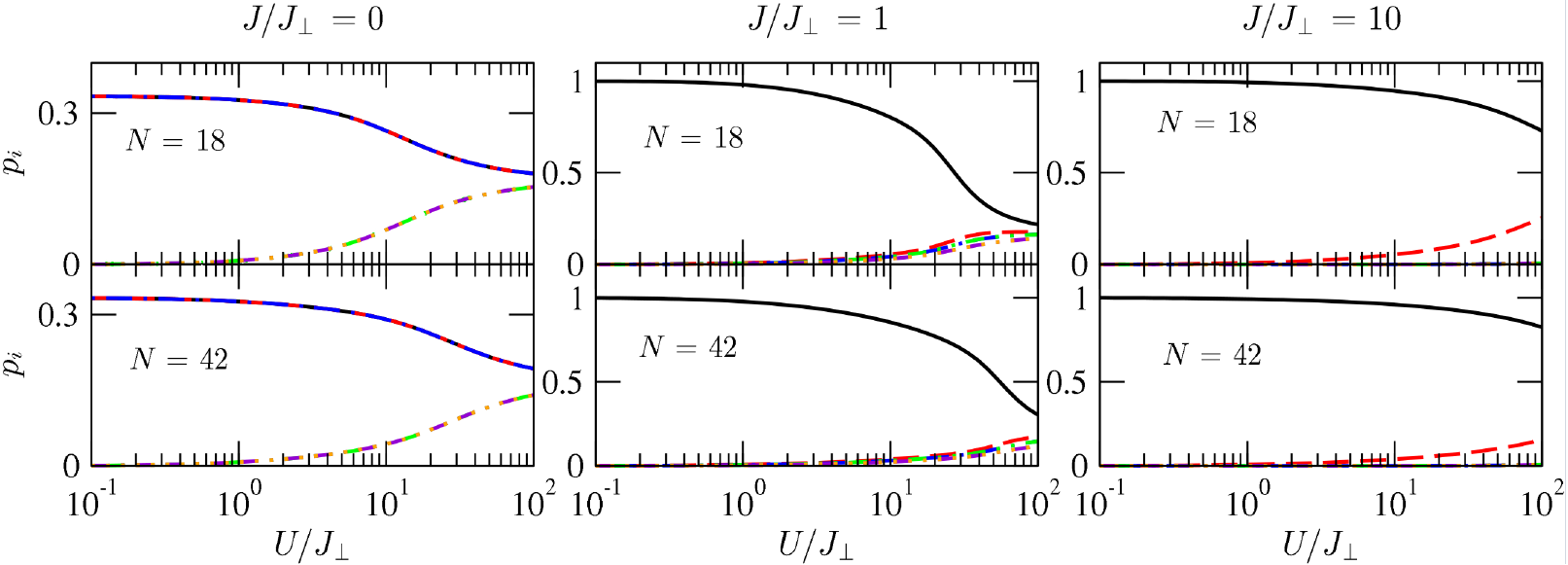} 
\caption{Eigenvalues of the one-body density matrix as a function of $U/J_\perp$ (in logarithmic scale) for different 
values of $J/J_{\perp}$, and $N$. 
Top panels correspond to $N=18$, and bottom ones to $N=42$.
Left panels correspond to $J/J_\perp=0$, middle panels to $J/J_\perp=1$ and right panels to $J/J_\perp=10$. 
Black solid line represents the largest eigenvalue $p_{1}$. 
The other eigenvalues $p_{i}$ are shown with dashed colored lines: 
red $p_{2}$, blue $p_{3}$, green $p_{4}$, violet $p_{5}$ and orange $p_{6}$. }
\label{eigenvalues_density_matrix_moreparticles}
\end{figure*}

We have extended the analysis of the condensation and fragmentation properties for
larger and commensurate number of particles confined in the two coupled rings. 
The results are summarized in 
Figs.~\ref{VNentropy_extra} and \ref{eigenvalues_density_matrix_moreparticles},
where we plot the von Neumann entropy and the eigenvalues of the one-body density matrix, respectively, as a 
function of $U/J_\perp$ for $N=18, 42$ atoms, and $M=3$.
In general, the same physical behavior is obtained for all the cases we have studied. In 
Fig.~\ref{VNentropy_extra} one can see that 
the limiting values of $S_{\rm VN}$ do not depend on the number of atoms confined in the two connected rings.
As in the previous case with $N=6$ atoms, independently of the number of commensurate atoms,
for small interactions the system is three-fragmented in three disconnected double-wells 
when $J/J_{\perp}=0$, whereas for $J/J_\perp \neq 0$ the system is condensed. In the other limit, for
large interactions the system tends to the Mott insulator regime more rapidly when $J/J_\perp \leq 1$.
As shown in Fig.~\ref{VNentropy_extra}, this asymptotic Mott insulator limit, $S_{\rm VN}=\ln(2M)$, 
when $J/J_\perp > 1$ is reached for larger values of $U/J_\perp$
when the number of atoms increases. Similarly, for a given ratio $J/J_\perp$, upon increasing the (integer) filling $\nu$, the behaviour of the Von Neumann Entropy is qualitatively unchanged, although the asymptotic value $S_{\rm VN}=\ln(2M)=\ln 6$ is reached for \textit{larger} values of $U$. This property clearly emerges from the comparison of Fig. \ref{VNentropy} and both panels of Fig. \ref{VNentropy_extra}.

The eigenvalues of the one-body density matrix for $N=18$ (top panels), 
and $42$ (bottom panels) are plotted in 
Fig.~\ref{eigenvalues_density_matrix_moreparticles} 
for different values of the tunneling ratios:
$J/J_\perp=0$ (left panels), $J/J_\perp=1$ (middle vertical panels) and $J/J_\perp=10$ (right panels).
When $J/J_\perp=0$, since the sites in the same ring are decoupled, the system behaves as three independent
double-wells and therefore $p_{1}=p_{2}=p_{3}$ and $p_{4}=p_{5}=p_{6}$ for any value of 
$U/J_{\perp}$.
For small interactions the system is condensed in the double ring $(p_1 \simeq 1)$ when $J/J_\perp \neq 0$,
or in three double-wells $(p_1=p_2=p_3=1/3)$ when $J/J_\perp=0$.
As before, for large interactions the system tends to the Mott insulator regime $p_i=1/(2M)$ $(i=1, 2,\dots,6)$.
This limit is achieved for smaller interactions $U/J_\perp$, when $J/J_\perp \leq 1$ for a fixed number of atoms,
or for smaller number of commensurate atoms when the tunneling ratio is fixed.

\subsubsection{Incommensurate number of atoms}

We have also investigated the coherence and fragmentation properties of the system when 
an incommensurate number of atoms (with respect to the total number of sites $2M$)
is trapped in the double ring.
Starting from the commensurate situation with $N$ atoms, we have considered the case with one added (or subtracted) extra particle, $(N+1)$ and $(N-1)$ atoms, respectively. 

\begin{figure}[t]
\includegraphics[width=1\linewidth]{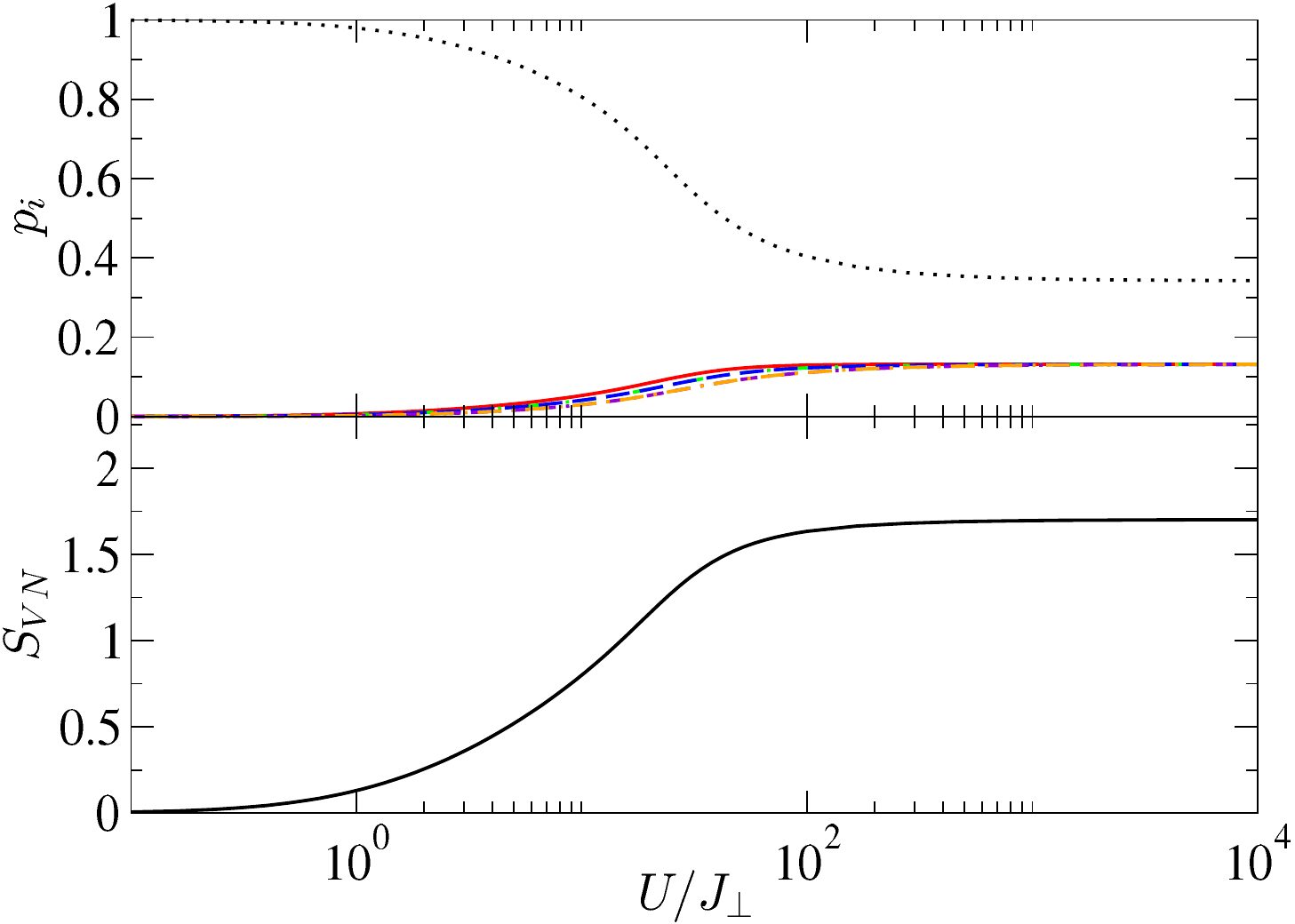} 
\caption{Eigenvalues of the one-body density matrix (upper panel) and von Neumann entropy (lower panel), as a function of $U/J_{\perp}$ for
$J/J_{\perp}=1$ computed with 19 atoms and $M=3$. 
Top panel: black dotted line represents the largest eigenvalue $p_1$, the other colored lines represent the other eigenvalues.}
\label{eigenvalues19}
\end{figure}

Figure \ref{eigenvalues19} shows the eigenvalues of the one-body density matrix (top panel) and its corresponding von Neumann entropy (bottom panel), as a function of $U/J_\perp$, 
obtained numerically by solving the Hamiltonian (\ref{hamiltonian})
for  two-connected rings, with $M=3$
and 19 atoms. This situation corresponds to adding one extra particle from the $N=18$ commensurate case 
studied previously with filling factor $\nu=3$.
For small interactions, $U/J_\perp < 1$, 
the ground state properties are almost the same with $N$ or $N+1$ atoms, and
the ground state behaves as in the commensurate case: $S_{\rm VN} \simeq 0$,
$p_1 \simeq 1$, and $p_j \simeq 0$ for $j \neq 1$. 
However, for large interactions $U/J_\perp > 10^2$,
when one atom is added, the system is not fully fragmented as in the commensurate
situation where the atoms are localized in the sites with filling factor $\nu=N/(2M)$. 
Here, the extra atom is not localized since it is 
shared with all the sites. 
Therefore the system cannot reach the fully fragmentation limit 
corresponding to the Mott insulator values $p_i=1/(2M)=1/6$ for all $i$, and $S_{\rm VM}=\ln(2M)=\ln 6$.

In the large interaction limit, 
when one atom is added (subtracted) from the commensurate case,
the analytical expressions of the corresponding one-body density matrix, $\hat{\rho}^{+}$ ($\hat{\rho}{-}$), 
and their eigenvalues, can be obtained from Eqs.~(\ref{wf-MIextra}) and (\ref{one-body}).
The diagonal matrix elements are given by $\rho^+_{(i,j)(i,l)}=\rho^-_{(i,j)(i,l)}=1/(2M)$,
and the off-diagonal elements are
$\rho^+_{(i,j)(k,l)}=(1/N)(N-1+2M)/(2M)^2$ and
$\rho^-_{(i,j)(k,l)}=(1/N)(N+1)/(2M)^2$.
From them it follows the whole set of eigenvalues: the largest one is
 $p^+_{1}=\rho^+_{(i,j)(i,l)}+(2M-1)\rho^+_{(i,j)(k,l)}$, and
 the others are
$p^+_{j}=\rho^+_{(i,j)(i,l)}-\rho^+_{(i,j)(k,l)}$.
The same expressions hold for the case of one subtracted atom, by
changing the components $\rho^+_{(i,j)(k,l)}$ by $\rho^-_{(i,j)(k,l)}$. 
These equations can be simplified by introducing the corresponding filling factor, 
which for one added extra particle is $\nu^+=\nu+1/(2M)$,
\begin{equation}
p^+_{1}=\frac{4\nu M+2M-\nu}{4\nu M^2+2M} \,, \quad p^+_{j}=\frac{2\nu M-\nu}{4 \nu M^2+2M} \,.
\label{m+}
\end{equation}
It is straightforward to prove that these expressions verify the eigenvalues relation of the one-body density matrix:
$\sum_{i=1}^{2M} p^+_i=p^+_{1}+(2M-1)\,p^+_{j}=1$. 
Analogously for the case of one subtracted particle, the filling factor is $\nu^-=\nu-1/(2M)$ and
\begin{equation}
p^-_{1}=\frac{4 \nu M-\nu-1}{4 \nu M^2-2M}\,, \quad p^-_{j}=\frac{2 \nu M-\nu-1}{4 \nu M^2-2M} \,.
\label{m-}
\end{equation}

In the limit of large interactions $U/J_\perp \gg 1$, 
the von Neumann entropy (\ref{eq-S_VN}) can be also computed analytically
from the corresponding eigenvalues of the one-body density matrix, Eqs.~(\ref{m+}) or (\ref{m-}). 
In both cases $S^+_{\rm VN}=S^-_{\rm VN}$, since adding or subtracting one particle, i.e. having 
one extra particle or one vacancy, is equivalent from the entropic point of view.
In the particular case of Fig.~\ref{eigenvalues19}, with $\nu=3$ and $18+1$ atoms, the limiting values for large interactions
computed analytically from Eqs.~(\ref{m+}) are:
$p^+_{1}=13/38$, $p^+_{j,}=5/38$ for $j \neq 1$, and  $S^+_{\rm VN}=1.726$,
which are in agreement with the numerical results calculated from the Hamiltonian.
In the incommensurate case with $18-1$ atoms, which corresponds to one subtracted atom
from the commensurate system $N=18$, the corresponding values 
in the limit of large interactions are 
$p^-_{1}=16/51$, $p^-_{j}=7/51$, but the same entropy  $S^-_{\rm VN}=1.726$ as adding one atom.

\subsection{Correlations}
\label{ss:co}
In general, entropy-based indicators are rather versatile quantities allowing for the detection of quantum phase transitions in one-component \cite{Bruno1} and multi-component \cite{Andrea_Entropy,Andrea_JPC,Andrea_NJP,Andrea_PRA3,Andrea_Zenesini} Bose-Hubbard models. Here, we focus on the correlations between particles on different sites, and we quantify it by means of the entropic indicator \cite{Bruno1}
\begin{equation}
S_{\rm C} = -\sum_{\alpha} |c_{\alpha} |^2 \ln |c_{\alpha} |^2 \,,
\end{equation}
where the term $c_\alpha$ is the coefficient of the Fock state $|\alpha \rangle$ in the expansion $\Ket{\Psi}=\sum_{\alpha} c_{\alpha} \Ket{\alpha}$. In essence, indicator $S_\mathrm{C}$ quantifies the degree of clustering of particles in the Fock space. The collection of all coefficients $|c_\alpha|^2$, in fact, can be regarded as a probability distribution over the space of Fock states (notice, also, that it is correctly normalized to $1$, since $\sum_\alpha|c_\alpha|^2=1$). 

%


If only one Fock state, $|\beta \rangle$, enters in the linear combination above, as in the case of a fully \textit{fragmented} state~\cite{condensed_fractions}, then the probability distribution is maximally peaked ($|c_\beta|^2=1$, $|c_\alpha|^2=0, \forall \alpha\neq\beta$), hence $S_{\rm C}=0$. Conversely, if two or more Fock states have non-zero projection on $|\Psi\rangle$, then the probability distribution $\{|c_\alpha|^2\}$ is broader and the associated entropy $S_{\rm C}$ is non-zero. For example, when the 
state of the system with 
$N=2M$ atoms forms a single Bose-Einstein condensate, Eq.~(\ref{wf-BEC}), 
the $S_{\rm C}$ can be calculated analytically as
$S_{\rm C} = N\ln(2M)-(1/2M)^{N}
\sum_{n_{1},...,n_{2M}}^{N}\binom{N}{n_{1},...,n_{2M}}\ln\binom{N}{n_{1},...,n_{2M}}$.

\begin{figure}[t]
\includegraphics[width=1\linewidth]{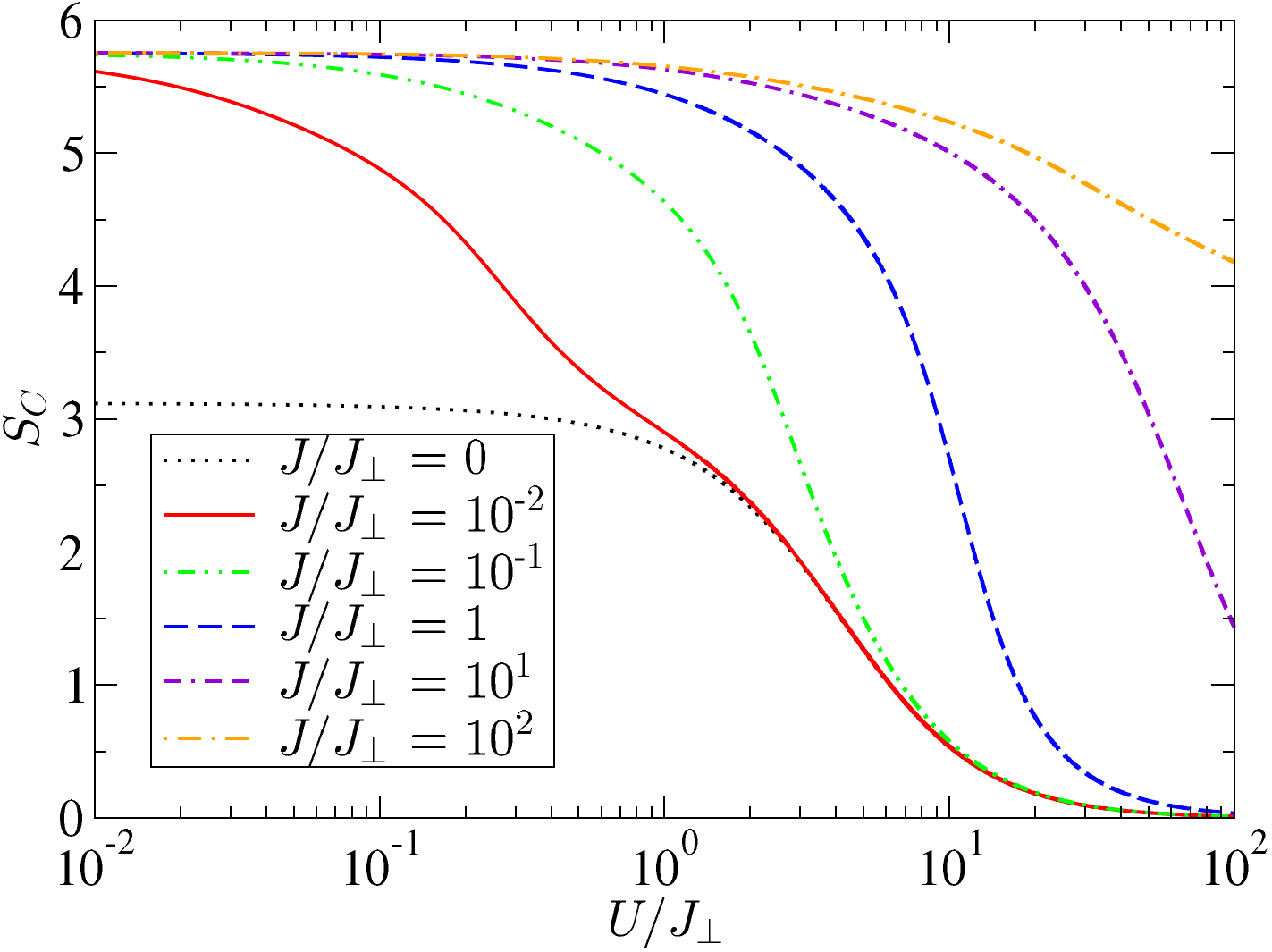} 
\caption{ $S_{\rm C}$ of the ground state as a function of $U/J_{\perp}$ 
for different values of $J/J_{\perp}$. In all cases, $N=6$ and $M=3$.}
\label{entropy_correlation}
\end{figure}

In Fig.~\ref{entropy_correlation} we plot $S_{\rm C}$ of the ground state of 
the two-coupled trimers, as a function of $U/J_\perp$ for the same values of 
$J/J_\perp$ as in Figs.~\ref{eigenvalues_density_matrix} and \ref{VNentropy}.
As we have already seen in the previous figure, when $U \ll J, J_\perp$ the 
system is almost condensed, but  $S_{\rm C}$ presents two limiting 
values: $S_{\rm C}=3.12$ when $J/J_{\perp}=0$, and $S_{\rm C}=5.76$ when 
$J/J_\perp \neq 0$. The latter value is in agreement with the previous analytical 
expression of $S_{\rm C}$ with $N=2M=6$. It means that when $U/J_\perp <1$ and 
$J/J_\perp \neq 0$ the ground state corresponds to one Bose-Einstein 
condensate. Moreover, the system remains condensed for larger values of the 
interaction $U/J_\perp$ as the tunneling ratio $J/J_\perp$ increases. Another 
physical situation appears in the particular case
when $J/J_{\perp}=0$; the system is decoupled into three
independent bosonic Josephson junctions with tunneling rate $J_\perp$ and 
$N/M=2$ bosons in each one. In this case, when the ratio $U/J_\perp < 1$, the bosons are 
occupying the same single-particle state of each Josephson junction. The 
entropy can be also computed analytically from the three independent 
double wells which yields $S_{\rm C}=3.12$ in agreement with our numerical results.

\begin{figure}[t]
\includegraphics[width=1\linewidth]{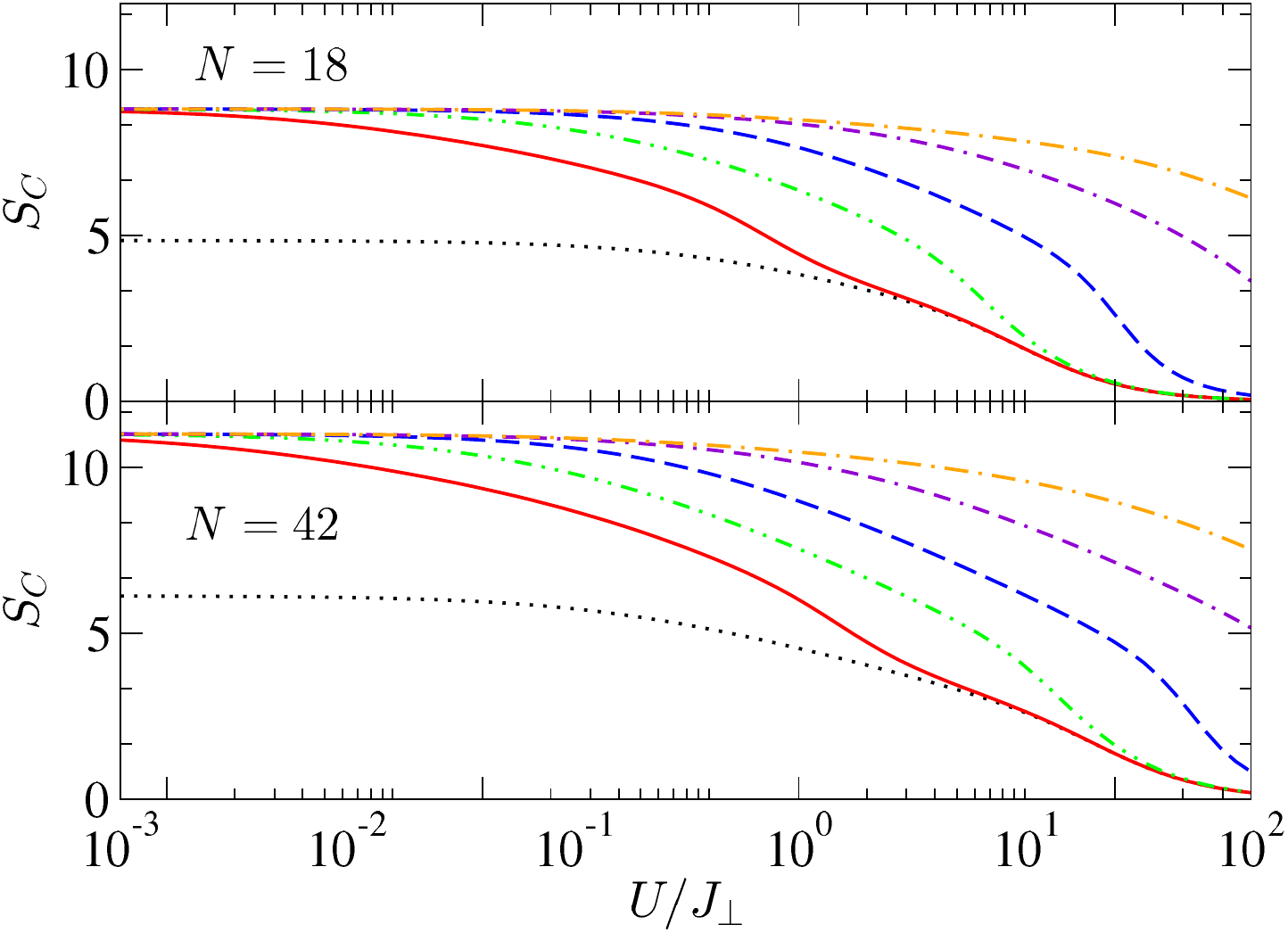} 
\caption{$S_{C}$ of the ground state as a function of $U/J_{\perp}$ 
for different values of $J/J_{\perp}$ and number of particles: $N=18$ (top panel),
and $N=42$ (bottom panel).  In all cases $M=3$.
The color legend is the same as in Fig.~\ref{entropy_correlation}.}
\label{entropy_correlation-extra}
\end{figure}

In the limit of large interactions $S_{\rm C}$ is given by 
$S_{\rm C} \rightarrow 0$ for all values of $J/J_\perp$; since 
in this situation the state of the system is fragmented and can be approximated by 
one Fock state with one particle localized in each site (Mott insulator 
regime). The largest localization with $S_{\rm C}=0$ is achieved for smaller 
values of the interaction $U/J_\perp$  as the tunneling ratio $J/J_\perp$ 
decreases. It is important to stress that if the number of particles is 
not commensurate with the total number of sites 
this yields $S_{\rm C} \neq 0 $ in the Mott limit, since the 
ground state is not given by only one Fock state but by a superposition of 
Fock states obtained by adding the extra particles that are shared with all 
sites~\cite{Wu2006}.
In the particular case of one added (subtracted) atom from a commensurate system,
$S_C^+$ $(S^-_{C})$ in the large interaction (Mott) regime can be
computed analytically using Eq.~(\ref{wf-MIextra}). 
For a given number of sites $M$,
$S^+_{C}=S^-_{C}=\ln(1/2M)$.

From Fig.~\ref{entropy_correlation} one can see that for a fixed value of 
$U/J_{\perp}$, the spatial correlation given by $S_{\rm C}$, increases for larger 
values of $J/J_{\perp}>1$ when the intra-ring coupling dominates over the 
inter-ring tunneling. This is in agreement with the results obtained from the 
previous quantities which indicate that the condensate fraction $p_1$ increases, 
and $p_1+p_2 \simeq 1$, when $J/J_{\perp}>1$ increases. In this case the 
distribution of non-zero Fock coefficients of the ground state is 
larger and thus the spatial correlations. 
Concerning the behaviour of $S_{\rm C }$ in the large-$N$ limit, comparing Fig. \ref{entropy_correlation} with both panels of Fig. \ref{entropy_correlation-extra}, one can observe that, for a given value of $J/J_\perp$, the asymptotic limit $S_{\rm C}=0$ is approached for a larger value of $U/J_\perp$ if the (commensurate) number of bosons $N$ is larger.

To get further insight into the correlations of the particles one can compute the pair correlation function between two particular sites of the system. In general, 
the pair correlation between site $i$ (of ring $j$) and site $k$ of (ring $l$) 
is defined as,
\begin{equation}
\eta_{(i,j),(k,l)} = 
\frac{\bra{\Psi} \hat{a}^{\dagger}_{i,j} \hat{a}^{\dagger}_{k,l} \hat{a}_{i,j} \,\hat{a}_{k,l} 
\Ket{\Psi}}{\rho_{(i,j),(i,j)}\,\rho_{(k,l),(k,l)}} \,.
\label{2pair-correlation}
\end{equation}
The value of $\eta_{(i,j),(k,l)}$ gives the probability of finding one particle 
in site $i$ (of ring $j$) when there is already a particle in site $k$ (of 
ring $l$). It is possible to write it in matrix form whose dimension is 
$2M \times 2M$, which contains all the pair correlations between all the sites.
In our system composed by two stacked identical rings, one can distinguish two types of 
pair correlations: intra-ring correlations $\eta^{1R}\equiv \eta_{(1,j),(2,j)}$ 
(correlations between two neighbor sites in the same ring coupled by $J$), 
and inter-ring correlations $\eta^{2R} \equiv \eta_{(i,\uparrow),(i,\downarrow)}$ 
(correlations between a pair of sites, connected by $J_\perp$, belonging to the two 
different rings).

In the two limiting regimes, the pair correlations 
can be computed also analytically. Interestingly, in each regime the intra-ring 
and the inter-ring pair correlations are equal. That is, the correlations between
two sites are the same independently of the position of the two sites considered.
In the superfluid regime, the ground state is condensed (\ref{wf-BEC}) 
and it yields $\eta_{\rm BEC}= (N-1)/N$. The pair correlation is independent of 
the number of sites since the atoms are completely delocalized. In the Mott insulator 
regime, the ground state has $\nu$ localized atoms in each site ($\nu \in \mathbb{Z}$) 
and from (\ref{wf-MI}) it follows $\eta_{\rm MI}=1$. This means that if there is 
one atom in one site, the probability to find an atom in another site is one.

\begin{figure}[t]
\includegraphics[width=1\linewidth]{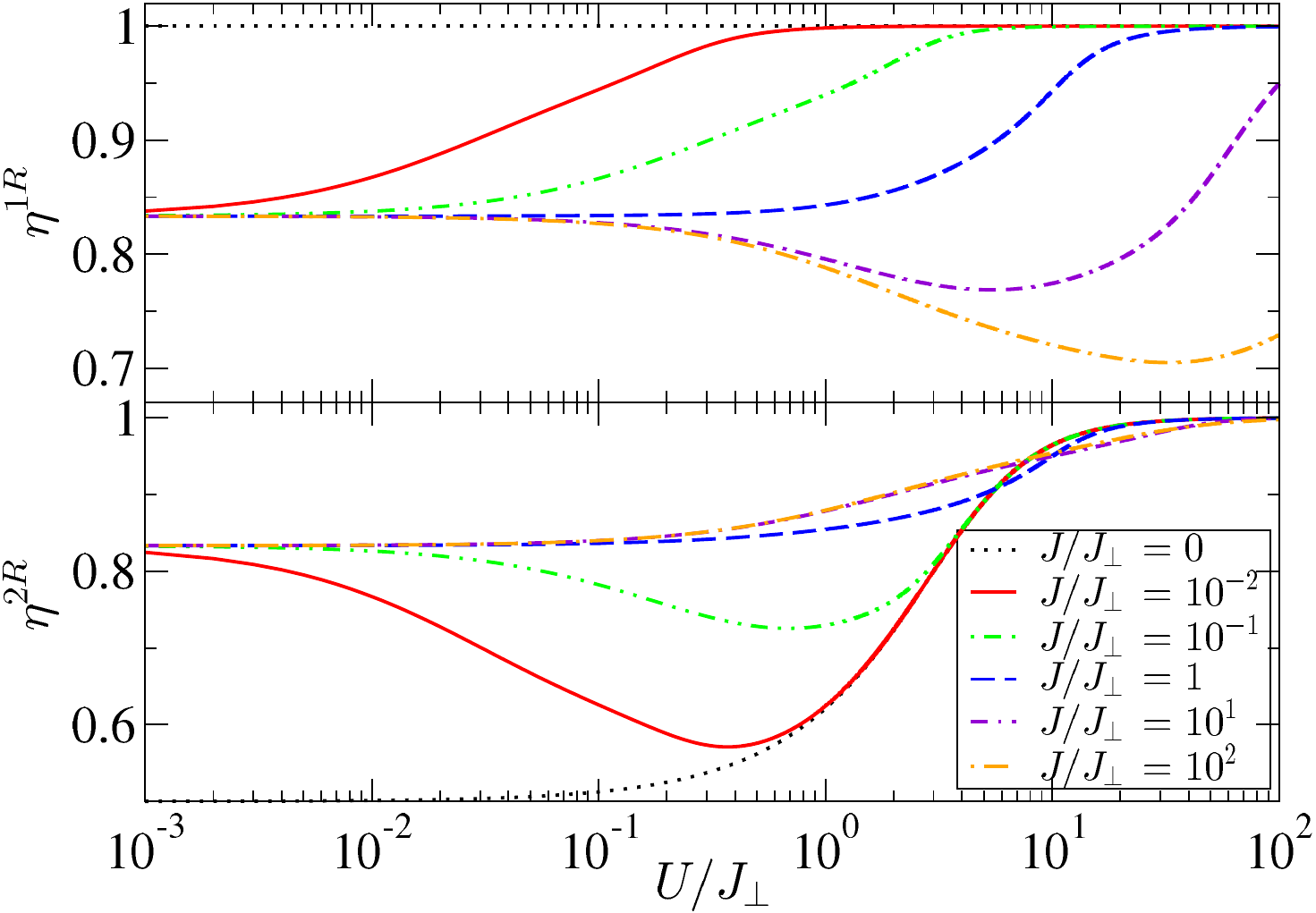} 
\caption{Pair intra-ring correlation (top) and inter-ring correlation (bottom), 
as a function of $U/J_{\perp}$ 
for different values of $J/J_{\perp}$. In all cases, $N=6$ and $M=3$.}
\label{paircorrelationinner}
\end{figure}
\begin{figure}[t]
\vspace{30pt}
\includegraphics[width=1\linewidth]{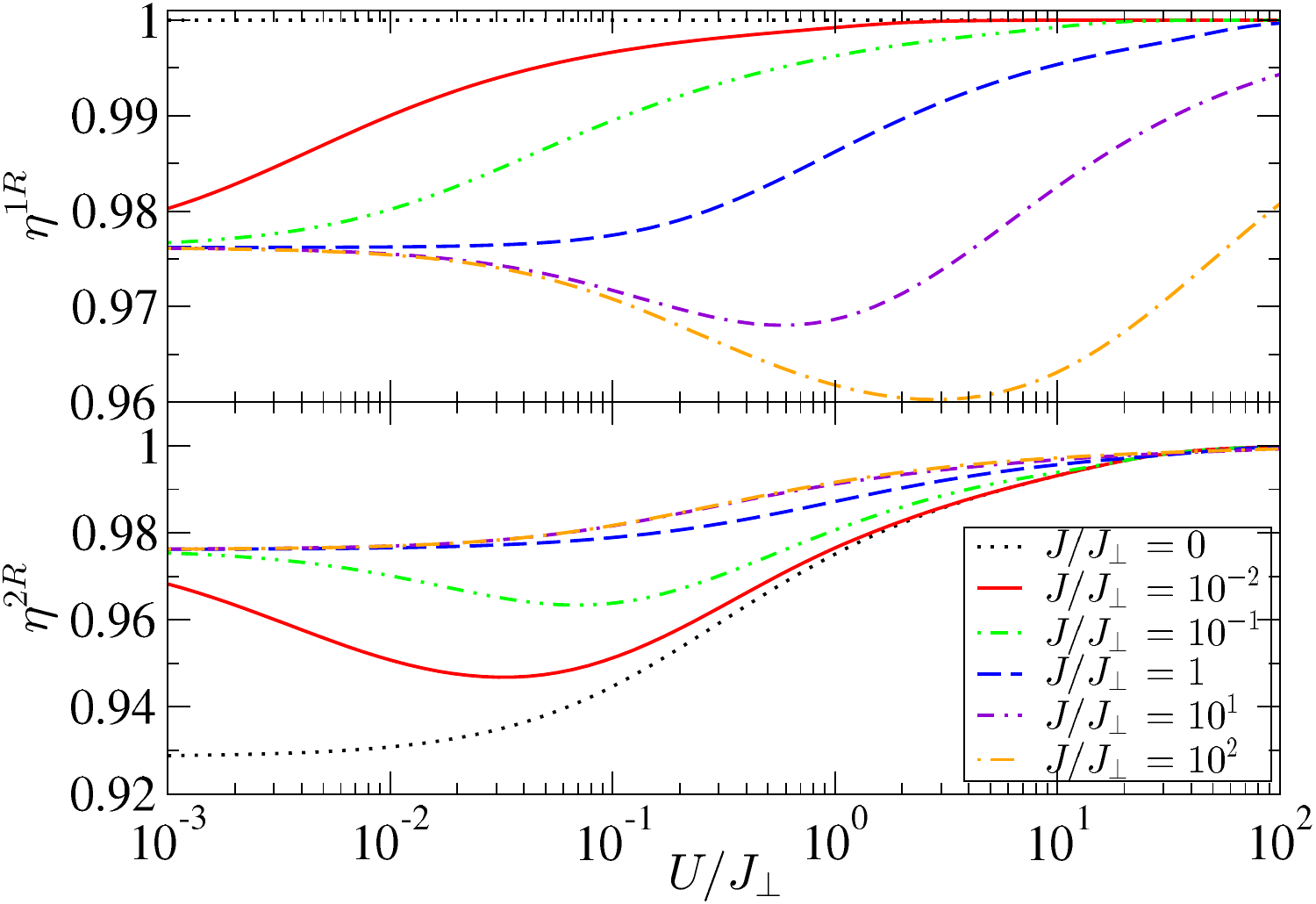} 
\caption{Pair intra-ring correlation (top) and inter-ring correlation (bottom), 
as a function of $U/J_{\perp}$ 
for different values of $J/J_{\perp}$. In all cases, $N=42$ and $M=3$.}
\label{paircorrelationinner-42}
\end{figure}

We have calculated numerically the pair correlations for two coupled trimers 
with $N=6$ (filling factor $1$), as a function of $U/J_\perp$ for different values 
of $J/J_{\perp}$. Our numerical results are presented in Fig.~\ref{paircorrelationinner}. 
The top (bottom) panel corresponds to the intra-ring (inter-ring) pair
correlations, $\eta^{1R}$ and $\eta^{2R}$, respectively. As expected, for small 
values of the interaction the system is coherent and the intra-ring and 
inter-ring pair correlations tend both to the superfluid limit $(N-1)/N=5/6$,
for all values of $J/J_\perp$ except for the particular case $J/J_\perp=0$. 

In the top panel of Fig.~\ref{paircorrelationinner} the pair intra-ring 
correlation between two adjacent sites of the same ring coupled by $J$ is 
depicted. A different behavior is appreciated when $J<J_\perp$ or $J>J_\perp$. 
When $J/J_\perp=0$ the two sites in the same ring are decoupled and the system 
behaves as three independent bosonic Josephson junctions. Since the two sites 
are independent, they become uncorrelated and $\eta^{1R}=1$ for all values 
of $U/J_\perp$. In the other cases, when  $J/J_\perp \neq 0$ and the interaction 
is small, the system is condensed and $\eta^{1R} \rightarrow (N-1)/N$.
As we have seen previously from the other magnitudes the system remains 
condensed for larger values of the interaction when $J$ increases, in 
particular when $J>J_\perp$. When $J/J_\perp \leq 1$ the intra-ring pair 
correlation has a smooth  and monotonous behavior from the coherent limit 
$\eta_{\rm BEC}=(N-1)/N=5/6$ to the Mott insulator value $
\eta_{\rm MI}=1$. Whereas for $J/J_\perp > 1$ the 
system behaves as two weakly coupled rings and the competition between tunneling 
strengths causes the appearance of a minimum in the intra-ring pair correlation.

The inter-ring pair correlation $\eta^{2R}$ is shown in the bottom panel 
of Fig.~\ref{paircorrelationinner}. Here we investigate the correlation between 
two sites that belong to different rings but are coupled by the tunneling 
strength $J_\perp$. The two limiting regimes appear clearly for all values 
$J/J_\perp \neq 0 \,$: the system remains coherent for small interactions 
($\eta^{2R} \to 5/6$) and tends to fragmentation for large interactions 
($\eta^{2R} \to 1$). In the particular case when $J/J_\perp =0$, the system acts 
as three independent double-wells, and the two sites behave as a bosonic Josephson 
junction, with $N=2$ particles and tunneling rate $J_\perp$, independently of 
the other sites; in this case when the interaction is small $(U/J_\perp < 10^{-1})$,
$\eta^{2R} =1/2$, and increasing the interaction 
the system goes smoothly to the Mott insulator value $\eta_{\rm MI}=1$.
The behavior of $\eta^{2R}$ increases for all values of the tunneling ratio
except when $J/J_\perp <1$, where a minimum appears for interactions $0.1< U/J_\perp<1$. 

Notice that when $J/J_\perp=1$ the intra- and inter-ring couplings are the same, 
and there is no physical difference between two connected sites in the same 
ring or connected in different rings. This leads to the same curve with 
$J/J_\perp=1$ for $\eta^{1R}$ and $\eta^{2R}$ in both panels of 
Fig.~\ref{paircorrelationinner}.

We show in Fig.~\ref{paircorrelationinner-42} the intra-ring and inter-ring pair correlations
for $N=42$ and $M=3$, as a function of $U/J_\perp$ for the same values of $J/J_\perp$ as in 
Fig.~\ref{paircorrelationinner}. 
The numerical results tend to the limiting values obtained analytically for commensurate systems in 
both limits, $U/J_\perp \ll 1 $ and $U/J_\perp \gg 1$. When $J/J_\perp \neq 0$, and 
for large interactions (Mott insulator regime)
$\eta^{1R} = \eta^{2R}=1 $, and for small interactions $\eta^{1R} = \eta^{2R}=(N-1)/N=41/42 $.
Whereas when $J/J_\perp=0$, $\eta^{1R} =1$ for all interactions, and $\eta^{2R}=1$ for large interactions.
In this latter tunneling ratio, the system behaves as three uncoupled double-wells with $N/3=14$ atoms,
therefore in the small interaction limit $\eta^{2R}=(N/3-1)/(N/3)=13/14$ which is in agreement with the 
numerical results.

From Figs.~\ref{paircorrelationinner} and \ref{paircorrelationinner-42} one can see that
the general behavior of the pair correlations is similar for different commensurate number of atoms.
However, the range of variation of $\eta^{1R}$ and $\eta^{2R}$ decreases when the number of trapped 
atoms increases, approaching to 1. 
Moreover, when $J/J_\perp \ll 1$ and 
the number of atoms increases, the system needs smaller values of $U/J_\perp$ to reach
the non-interacting limit $\eta^{1R} = \eta^{2R}=(N-1)/N$ (see for instance 
the red curve $J/J_\perp =10^{-2}$).

The pair correlations in the double ring system change for large interactions in the 
Mott regime when we add (subtract) a particle from a commensurate situation. Inserting the 
corresponding incommensurate ground states (\ref{wf-MIextra}) in the pair
correlation definition, Eq.~(\ref{2pair-correlation}) it follows:
\begin{equation}
\eta^{\pm}_{(i,j),(k,l)}= 4M\frac{(M-1)\nu^2+\nu(\nu \pm 1)}{\left[(2M-1)\nu+(\nu \pm 1)\right]^2} \,,
\label{eq:eta_pm}
\end{equation}
where $\pm$ stands for adding (subtracting) an atom
from the commensurate case $\nu \in \mathbb{Z^+}$, respectively. 
It is straightforward to 
see that the intra-ring and inter-ring correlations ($\eta^{1R}$ and $\eta^{2R}$) 
are equal in this regime,
since the extra particle (or vacancy) is equally populating all the sites. Eventually, one can verify that, both in the commensurate and in the incommensurate case, in the limit of large number of bosons $N$, the pair correlations tend to $1$ for any value of $U/J_\perp$. This can be observed when comparing Figs.~\ref{paircorrelationinner} and \ref{paircorrelationinner-42} and can be easily understood by computing the large-$N$ limit of $\eta_{\rm BEC}=(N-1)/N$ and of Eq. (\ref{eq:eta_pm}).

\subsection{Reduced density matrix: Schmidt gap and entanglement entropy}
\label{ss:rdm}
\begin{figure}[t]
\includegraphics[width=0.8\linewidth]{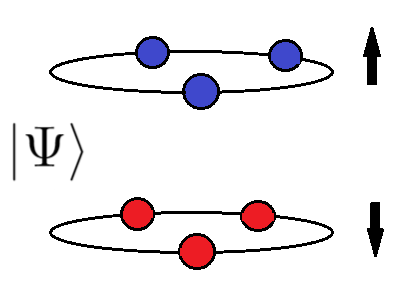} 
\caption{Sketch of the bipartite splitting of the system. Each ring acts as a subsystem. 
}
\label{fig:diagram}
\end{figure}

The spatial entanglement properties of the ground state $\Ket{\Psi}$ can be characterized from the features
of the reduced density matrix, obtained by performing a bipartite 
splitting of the system~\cite{DeChiara2018}. We consider the two rings as the natural partition in 
two subsystems in order to investigate the entanglement 
between them (see Fig.~\ref{fig:diagram} for a sketch of the bipartite splitting of the system). From the density matrix $\rho = \Ket{\Psi} \bra{\Psi}$, 
tracing out the degrees of freedom of one ring, for example the bottom 
one, we obtain the reduced density matrix on the other subsystem, 
${\rho}^\uparrow$, that describes the state of the top ring.
The eigenvalues of ${\rho}^\uparrow$ are the Schmidt coefficients 
$\lambda_k^{\uparrow}$, which satisfy $\lambda_1^{\uparrow} > \lambda_2^{\uparrow} >... \,$, 
and $\sum_{k} \lambda_{k}^{\uparrow}=1$. The coefficient $\lambda_k^{\uparrow}$ 
represents the probability of finding $k$ particles in the top ring 
without measuring the number of particles in the other ring~\cite{gallemi2}.

The Schmidt gap is defined as the difference between the two largest 
coefficients of the Schmidt spectrum: 
$\Delta \lambda^{\uparrow}=\lambda_1^{\uparrow}-\lambda_2^{\uparrow}$, and it 
quantifies the entanglement of the two subsystems~\cite{DeChiara2018}. In the case of two 
subsystems without entanglement the Schmidt gap is $\Delta \lambda^{\uparrow}=1$, 
and the state of the system can be written as a product state.
Whereas when $\Delta \lambda^{\uparrow}=0$ there is a large entanglement 
between the two subsystems and the 
total state of the system cannot be expressed as a product state.

An alternative description of the entanglement between the two rings (subsystems) 
can also be obtained from the single-ring von Neumann entropy,
defined as $S^\uparrow= -{\rm Tr} ({\rho}^\uparrow \log {\rho}^\uparrow)$. 
Using the eigenvalues of the reduced density matrix, we can rewrite
$S^{\uparrow} = -\sum_{k} \lambda_{k}^\uparrow \log {\lambda_{k}^\uparrow}$.
In the Mott insulator regime, since there is only one non-vanishing 
Schmidt coefficient $\lambda_{1}^\uparrow=1$, this leads to $S^{\uparrow}_{\rm MI}=0$.
In the other limit, when the system is superfluid, the state is not separable 
in the two rings and the entanglement entropy is non zero, $S^{\uparrow}_{\rm BEC} \neq 0$.
Tracing out one of the rings in the density matrix we obtain the following 
expression for the reduced density matrix $\rho^{\uparrow}$ (see equation 
(\ref{reduced-density}) in the appendix),
\begin{equation}
 \rho^{\uparrow} = \sum_{\alpha,\beta} \sum_{m}\delta_{n(\alpha),n(\beta)}
 C_{\alpha,m}C_{\beta,m}^{*}\Ket{\alpha}\bra{\beta} \,,
\end{equation}
where $n(\alpha)$ is the number of particles of the Fock state $\Ket{\alpha}$ 
in the top ring subspace, and $C_{\alpha,m}$ are the coefficients of the ground state 
in the Fock basis of the two rings system.
\begin{figure}[t]
\includegraphics[width=1\linewidth]{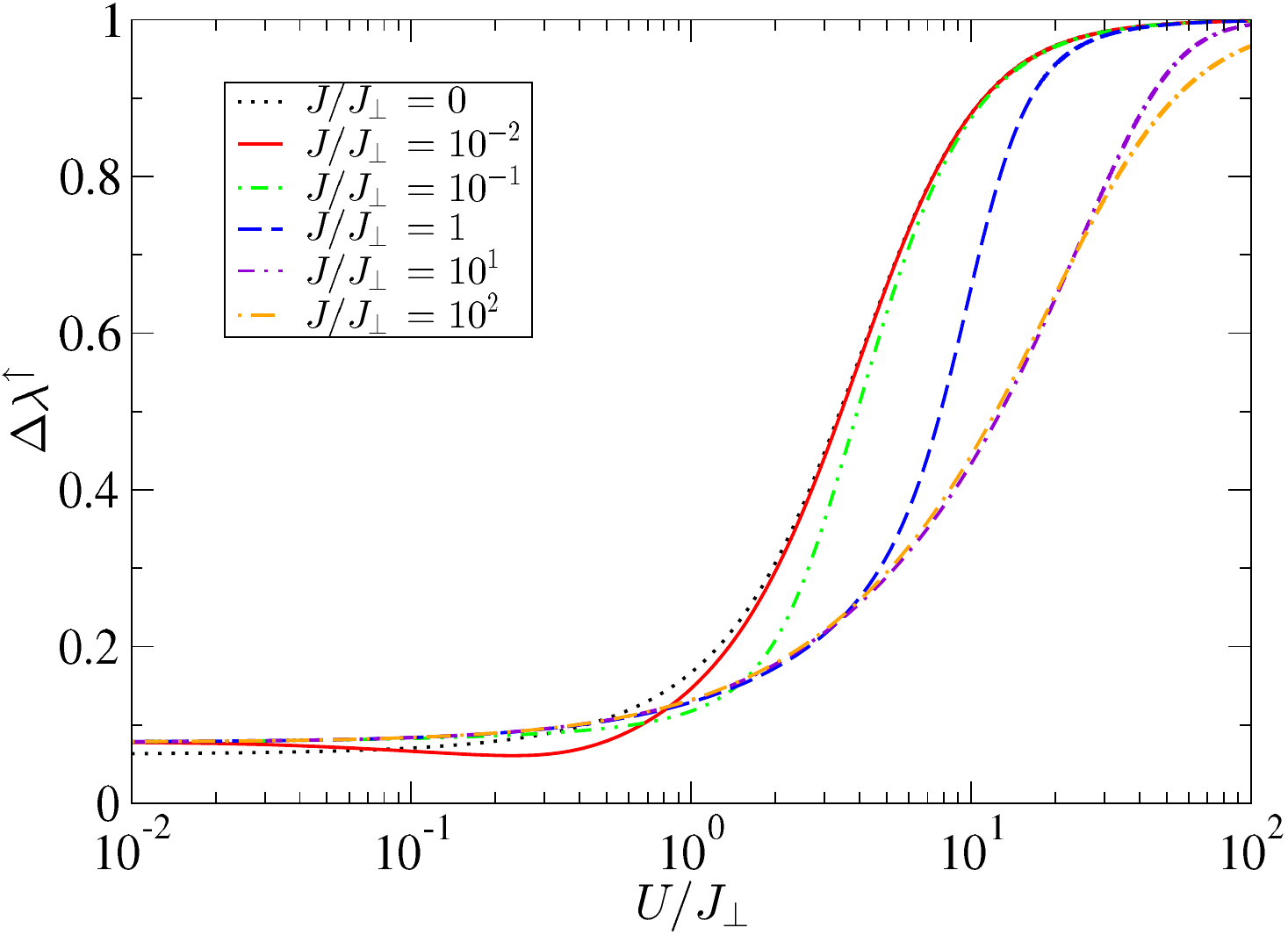} 
\caption{Schmidt gap of the reduced density matrix of the top ring as 
a function of $U/J_{\perp}$ for different values of $J/J_{\perp}$. In all cases, 
$N=6$ and $M=3$. 
}
\label{gap}
\end{figure}

\begin{figure}[t]
\includegraphics[width=1\linewidth]{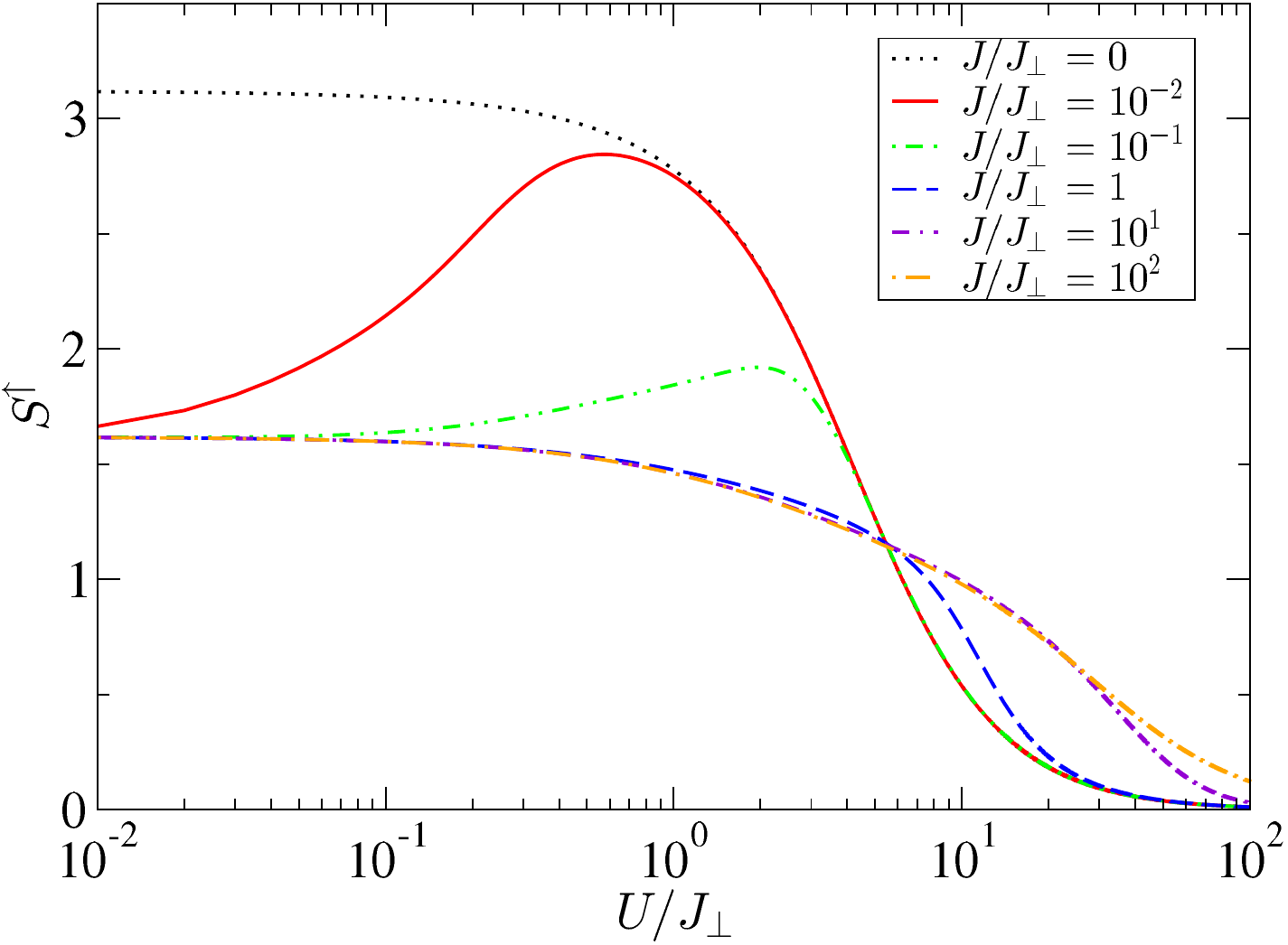} 
\caption{Entanglement entropy $S^{\uparrow}$ as a function of $U/J_{\perp}$ for 
different values of $J/J_{\perp}$. In all cases, $N=6$ and $M=3$. 
}
\label{entropy_gap}
\end{figure}

In Figs.~\ref{gap} and~\ref{entropy_gap} we plot the Schmidt gap 
and the entropy associated to the Schmidt coefficients of the subsystem formed by 
the top ring, with $N=6$ and $M=3$. The two limiting cases appear clearly in both figures: 
large tunneling couplings $U/J_{\perp} \ll 1$ (superfluid limit) and large interactions 
$U/J_{\perp} \gg 1$ (Mott insulator limit).
The Mott insulator regime ($\Delta \lambda^{\uparrow}=1$ and $S^\uparrow =0$) is reached 
for different values of the interaction strength $U/J_\perp$ depending on $J/J_\perp$. 
In this regime the state of the system is localized, with $\nu$ atoms in each site 
(when $\nu \in \mathbb{Z}$). Thus, it corresponds to only one Fock state that can 
be written as a product state of the top and bottom ring subspace.  Therefore the 
two rings are decoupled. Since they are not entangled, it means that $\Delta \lambda^{\uparrow}=1$.

In the other limit, when $U/J_\perp \rightarrow 0$ and $U/J \rightarrow 0$ 
(with $J/J_{\perp} \neq 0$), we have checked that the first $(N+1)$ Schmidt 
coefficients that we have obtained numerically are equal to 
$\lambda_{k}^{\uparrow}=2^{-N}\binom{N}{k}$, with $k=0,1,...,N$, that correspond 
to the Schmidt spectrum of a bosonic Josephson junction, see equation 
(\ref{binomial}) in the appendix, whereas the remaining Schmidt coefficients 
vanish. Thus, when the system approaches the coherent regime and $N$ is 
even\footnote{Note that in the superfluid regime, due to the binomial distribution, 
the Schmidt gap will always vanish when $N$ is odd.}, the Schmidt gap is given by 
$\Delta \lambda^{\uparrow}_{\rm BEC}=2^{-N}\left[\binom{N}{N/2}-\binom{N}{N/2-1}\right]$.
In this regime the atoms are delocalized and therefore the probability to find 
a particle in one site is the same in all the sites, leading to a binomial 
distribution. In our case $\Delta \lambda^{\uparrow}= 5/2^6 \simeq 0.08$ as it appears 
in Fig.~\ref{gap} when $U/J_\perp  \ll1$. In this regime, when one ring is traced 
out, the system presents the same features as one bosonic Josephson junction.
Thus, each ring behaves as an effective single site, while the whole system 
acts as an effective bosonic Josephson junction between the two rings.

In Fig.~\ref{entropy_gap} we show the entanglement entropy of the top ring.
For large values of the interaction $U/J_\perp$ the system tends to the Mott 
insulator value $S^\uparrow \rightarrow 0$. On the contrary, in the coherent
regime when $J/J_\perp \neq 0$, $S^\uparrow \simeq 1.62$, which corresponds
to the entanglement entropy for one site in a bosonic Josephson junction (with $N=6$):
$S^{\uparrow}=N\ln(2)-2^{-N}\sum_{k=0}^{N}\binom{N}{k}\ln \binom{N}{k}$ (see appendix). 
This limit is not reached when $J/J_{\perp}=0$ since in this case the tunneling 
between the sites in the same ring vanishes and the system corresponds to $M=3$ 
uncoupled bosonic Josephson junctions. In this situation the 
ground state is the product state of three individual Josephson wave functions which 
yields $S^\uparrow \simeq 3.1$.

From the above entanglement results, it follows that the 
ground state of two-coupled discrete circuits, when the number of bosons is even, can only 
be written as a product state of the two rings separately  when the interaction 
is large in front of the tunneling (Mott insulator regime).

\begin{figure}[t]
\includegraphics[width=1\linewidth]{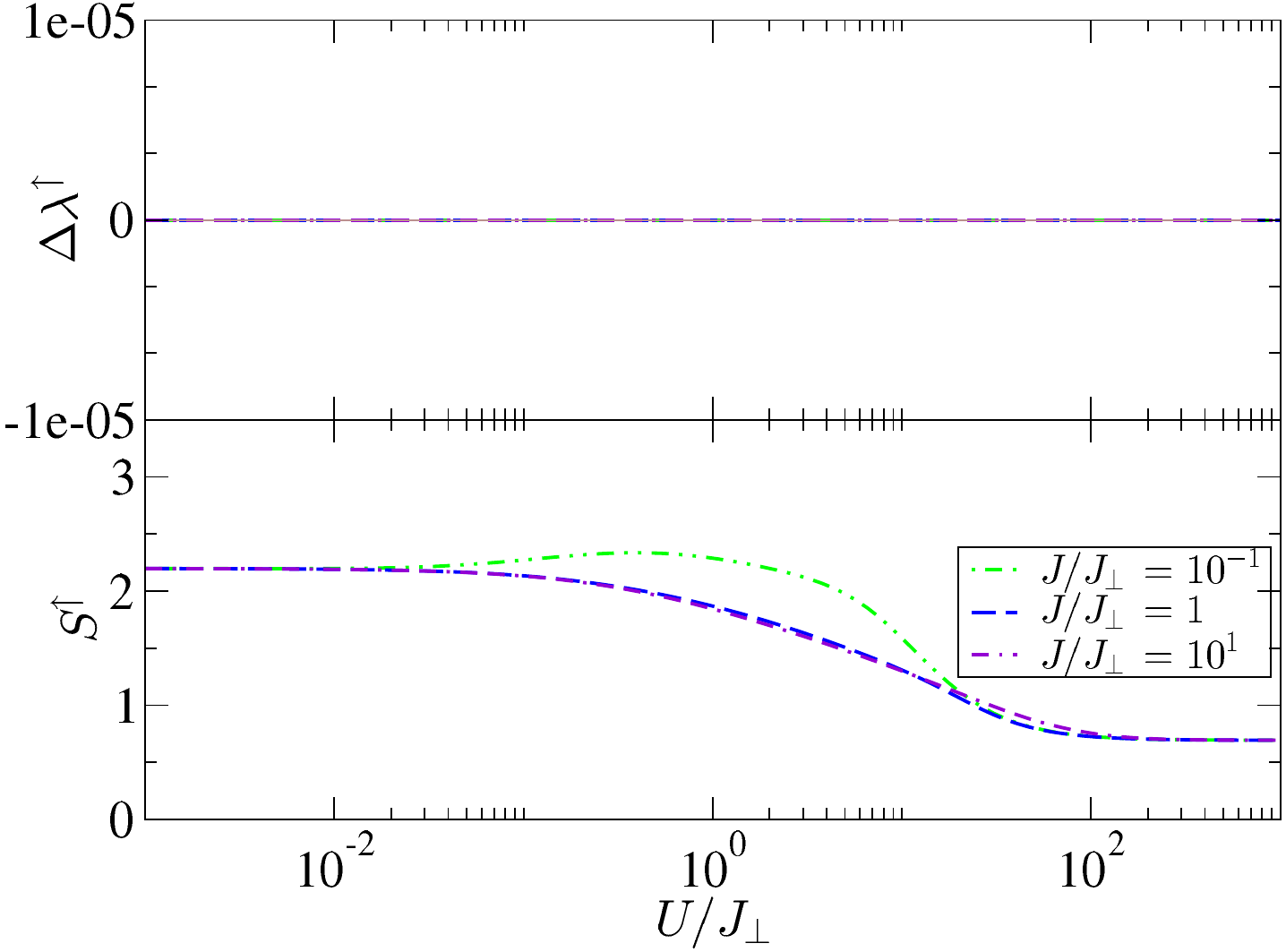} 
\caption{Schmidt gap $\Delta \lambda^\uparrow$ (top panel) and 
entanglement entropy $S^{\uparrow}$ (bottom panel) of the top ring, as a function of $U/J_{\perp}$ for 
different values of $J/J_{\perp}$. In all cases, $N=19$ and $M=3$.}
\label{cosica}
\end{figure}

Let us consider a non commensurate number of atoms in the double ring. 
Figure \ref{cosica} depicts the numerical calculations of the Schmidt gap (top panel)
and the Schmidt entropy (bottom panel) of the top ring for a system with $M=3$ and
19 atoms. It corresponds to adding an extra atom from the commensurate case with $N=18$.
In the limit of small interactions the system has the same physical behavior independently
of the number of trapped atoms; there is a macroscopic occupation of a single particle
state, and therefore $p_1 \simeq 0$. The state of the system cannot be expressed as a product
state of the two rings, then $\Delta \lambda^\uparrow \simeq 0$ and the two rings are entangled.
Notice that for an odd number of particles, which is incommensurate with $2M$ sites,
the Schmidt gap 
always vanishes for any value of the interaction $U/J_{\perp}$, and therefore the system is always 
entangled, as it shown in the top panel of Fig.~\ref{cosica}.

As we have already previously discussed, 
in general, for large interactions the physical quantities are 
more sensitive to the non commensurability:
the ground state is composed by a certain number of Fock states, due to the fact that
the extra particle is shared between all the sites, and 
the two rings become entangled. In particular, 
when the system has one added (subtracted) atom from a commensurate case,
the ground state can be described by Eq.~(\ref{wf-MIextra}).
In this situation  
the reduced density matrix and therefore the Schmidt coefficients can be analytically computed. 
It follows that there are
only two non-vanishing Schmidt coefficients $\lambda_{0}^\uparrow=\lambda_{1}^\uparrow=1/2$, 
which leads to $\Delta \lambda^\uparrow=0$ and $S^{\uparrow} = \ln(2)$. 
These limiting values are in agreement with the numerical results shown in Fig.~\ref{cosica}
for large interactions ($U/J_\perp > 10^2$).
As expected, this situation is 
clearly different from the commensurate one where 
$\Delta \lambda^\uparrow=1$ and $S^{\uparrow} = 0$ (see 
Figs.~\ref{gap} and~\ref{entropy_gap}).

Eventually, it is worth emphasizing that the entanglement entropy has proved to be a rather effective indicator in detecting the occurrence of the superfluid-Mott-insulator transition (see Ref. \cite{Roscilde_Entanglement,Bruno_Ivan} and references therein). In general, if one considers the well-known $(J/U,\mu/U)$ phase diagram of the Bose-Hubbard model (where $\mu$ is the chemical potential), it is possible to draw constant-filling lines, both in the commensurate, and in the incommensurate case. Moving from the weakly interacting to the strongly interacting regime along a \textit{commensurate} constant-density line, the tip of the $\nu$-th Mott lobe is crossed before entering in the lobe itself. Accordingly, the entanglement entropy is known to exhibit a characteristic behavior \cite{Roscilde_Entanglement}: it is non-zero in the superfluid region, maximum at the tip of the Mott lobe, and zero inside the latter. Indeed, this is exactly the scenario which emerges from the analysis of our two-ring system (see the red solid and the green dot-dashed lines in Fig. \ref{entropy_gap}). Conversely, if one moves from the weakly interacting to the strongly interacting regime along an \textit{incommensurate} constant-density line, no phase border is crossed. This is because such a line ends up being wedged between two adjacent Mott lobes in the $(J/U,\mu/U)$ plane, a circumstance corresponding to the fact that the superfluid solution is resilient to strong interactions. The entanglement entropy along this constant-density line is known to be a smooth function of the interaction and, as opposed to the commensurate case, it does \textit{not} tend to zero in the strongly interacting regime. This is exactly what we observe when we compute the entanglement entropy of our two-ring system in the \textit{incommensurate} case (see the bottom panel of Fig. \ref{cosica}). 

\section{Conclusions}
\label{sec6}

In this work, we have considered a bosonic two-ring ladder, a system which is of particular interest in the emergent field of Atomtronics \cite{Amico_Roadmap}, as it can be suitably engineered to work as a supercurrent-based qubit device, it allows to explore magnetic-like phases in the presence of artificial gauge fields, and it is prone to disclose a rather rich dynamical scenario where persistent currents and vortices can tunnel between the rings. In addition, the two-ring ladder system does interpolate between other two classes of prototypical atomtronic devices: the famous double-well geometry (also known as bosonic Josephson junction) and the (single) Bose-Hubbard ring. Since we explored extended ranges of model parameters, the presented results have captured not only standard "asymptotic limits", but also interesting cross-over regimes, where the competition among different couplings is at its most crucial.

Within a simple analytical framework and by means of the exact numerical diagonalization of the system's Hamiltonian, we have investigated the static properties of the ground state.
We have shown that the competition between the two tunneling strengths (intra-ring and inter-ring 
coupling), together with the interaction, determine the quantum properties of the two connected rings.
We have also discussed the effects of the non commensurability when one atom is added (subtracted) from
the commensurate situation, and we have shown that these effects become more important for large interactions.

First we have studied the single-particle ground state, which can be interpreted as different 
configurations of the same type of vortex states in both rings. 
Then we have considered the interacting system with different number of trapped atoms, 
providing a comprehensive analysis of the quantum features of the system. We have investigated the
coherence and fragmentation properties, where we have found that the 
tunneling parameter $J$ is dominant in comparison with $J_{\perp}$. 
We have  also studied the quantum correlations between sites. 
In general, the pair correlations increase with the interaction. However,
there is a range of values, $J/J_\perp >1$ ($J/J_\perp <1$), 
where there is a competition between tunneling strengths
that causes the appearance of a minimum in the  intra-ring (inter-ring) pair 
correlation for moderate values of the interaction.
We have studied the entanglement of the ground state between the 
two rings computing its reduced density matrix. We have pointed out that in 
the superfluid case the non-null Schmidt coefficients of the 
system are the same as of a bosonic Josephson junction, and that the system 
is always entangled when the number of trapped atoms is 
incommensurate with the total number of sites of the system. Thus, the ground state 
can only be written as a product state in the Mott-insulator 
regime and with an integer value of the filling factor. 

These results show a number of interesting quantum features such as correlations 
between two sites and quantum 
entanglement that can be used to design atomtronic circuits with two 
linearly coupled stacked rings. 

\begin{acknowledgments}
We thank useful discussions with Alessio Celi. 
A. E. also thanks Claudia Gonzalez, Jakub Janarek and Pere Mujal 
for interesting discussions. We acknowledge  financial support from the Spanish 
MINECO and the Fondo Europeo de Desarrollo  Regional (FEDER, EU) under
Grants No. FIS2017-87801-P and FIS2017-87534-P, 
and from Generalitat de Catalunya under Grant No.  2017SGR533.
A. E. is supported by Spanish MECD fellowship FPU15/03583.
\end{acknowledgments}

 \appendix
\begin{widetext}
\section{Reduced density matrix of one ring}
\label{app:reduced-density}
In order to obtain the set of eigenvalues $\{\lambda_k^\uparrow\}$ or Schmidt spectrum,
one has to obtain the reduced density matrix $\rho^\uparrow$ by tracing out the degrees of freedom of the other ring. 
We present the general case of two coupled rings with $M$ sites each one, and $N$ particles.
The dimension of the corresponding Hilbert space $\mathscr{H}$ is given by 
${\rm dim}(\mathscr{H}) =  \mathcal{N}_{N}^{2M}  =(N+2M-1)!/[N!(2M-1)!]$.
We consider the bipartite splitting in two subsystems which 
are given by the two rings $\mathscr{H_{\uparrow}}$ and $\mathscr{H_{\downarrow}}$. The dimension of this 
two subspaces is equal, given by ${\rm dim}(\mathscr{H_{\uparrow}}) = \sum_{n=0}^{N}  \mathcal{N}_{n}^{M}  = 
\frac{(N+1)}{M}\mathcal{N}_{N+1}^{M+N}$, but $\mathscr{H} \neq \mathscr{H_{\uparrow}} \otimes \mathscr{H_{\downarrow}}$.
The many-body ground state expressed in the Fock basis of the whole system reads,
\begin{equation}
\Ket{\Psi} = \sum_{n_{1},n_{2},...,n_{2M}} C_{n_{1},n_{2},...,n_{2M}} \Ket{n_{1},n_{2},...,n_{2M}},
\end{equation}
where $n_{k}$ are number of particles on site $k$. The first sites $k=1, ... M$ correspond
to the top ring, 
whereas $k=M+1,... 2M$ to the bottom one.
%
Since the total number of atoms is $N=\sum_{l=1}^{2M}n_{l}$, we can rewrite the many-body state using this constraint:
\begin{equation}
\Ket{\Psi} = \sum_{\lbrace n_{k} \rbrace} C_{n_{1},...,n_{M},n_{M+1},...,n_{2M-1}} \Ket{n_{1},...,n_{M},n_{M+1},...,n_{2M-1},N-\sum_{l=1}^{2M-1}n_{l}}\,.
\end{equation} 
The eigenvectors of the Fock basis can be expressed as a product of the Fock basis elements corresponding to the subspaces
of each ring:
\begin{equation}
 \Ket{n_{1},...,n_{M},n_{M+1},...,n_{2M}} = \Ket{n_{1},...,n_{M}}_{\uparrow}\otimes
 \Ket{n_{M+1},...,n_{2M-1},N-\sum_{l=1}^{2M-1}n_{l}}_{\downarrow} \,.
\end{equation}
The reduced density matrix of 
the top ring, $\rho^{\uparrow}$, is obtained by tracing out the degrees 
of freedom of the other subsystem:
\begin{align}
 \rho^{\uparrow} &= \sum_{\lbrace m_{k}\rbrace} \bra{m_{2M}}\otimes...\otimes\bra{m_{M+1}}\left( \Ket{\Psi}
 \bra{\Psi} \right)\Ket{m_{M+1}}\otimes...\otimes\Ket{m_{2M}}\,\nonumber\\
    &=\sum_{\lbrace m_{k}\rbrace}\sum_{\lbrace n_{k}\rbrace}\sum_{\lbrace n'_{k}\rbrace}
    \left(\prod_{l=M+1}^{2M-1}\delta_{n_{l},m_{l}}\delta_{n'_{l},m_{l}}\right) \delta_{N-\sum_{l=1}^{2M-1}n_{l},m_{2M}} 
    \delta_{N-\sum_{l=1}^{2M-1}n'_{l},m_{2M}} \nonumber \\
&\times C_{n_{1},...,n_{2M-1}}C_{n'_{1},...,n'_{2M-1}}^{*} \Ket{n_{1},...,n_{M}} 
    \bra{n'_{1},...,n'_{M}}\,\nonumber\\
    &=\sum_{\lbrace m_{k}\rbrace}\sum_{\lbrace n_{k}\rbrace}\sum_{\lbrace n'_{k}\rbrace} 
    \delta_{N-\sum_{l=1}^{M}n_{l}-\sum_{l=M+1}^{2M-1},m_{2M}} 
    \delta_{N-\sum_{l=1}^{M}n'_{l}-\sum_{l=M+1}^{2M-1},m_{2M}} \nonumber \\
&\times C_{n_{1},...,n_{M},m_{M+1},...,m_{2M-1}}C_{n'_{1},...,n'_{M},m_{M+1},...,m_{2M-1}}^{*} 
    \Ket{n_{1},...,n_{M}} \bra{n'_{1},...,n'_{M}}\,\nonumber\\
    &= \sum_{\lbrace m_{k}\rbrace}\sum_{\lbrace n_{k}\rbrace}\sum_{\lbrace n'_{k}\rbrace} \delta_{\sum_{l=1}^{M} n_{l}, 
    \sum_{l=1}^{M} n'_{l}} C_{n_{1},...,n_{M},m_{M+1},...,m_{2M-1}}C_{n'_{1},...,n'_{M},m_{M+1},...,m_{2M-1}}^{*} 
    \Ket{n_{1},...,n_{M}} \bra{n_{1},...,n_{M}} \,. 
\end{align}
Note that the condition $\delta_{\sum_{l=1}^{M} n_{l}, \sum_{l=1}^{M} n'_{l}}$ in the previous expression is crucial, 
since only Fock states $\Ket{n_{1},...,n_{M}}$ and $\bra{n'_{1},...,n'_{M}}$ that have
the same number of particles, i.e, 
$\sum_{l=1}^{M} n_{l}=\sum_{l=1}^{M} n'_{l}$, will contribute. 
This will be very useful in order to perform numerical computations. 

The following steep is to diagonalize $\rho^{\uparrow}$ in order to obtain the eigenvalues 
$\lbrace \lambda_{i}^{\uparrow} \rbrace$ and from the two largest ones compute the Schmidt gap. 
In general, for any number of sites $M$ and particles $N$ the matrix $\rho^{\uparrow}$ is given by:
\begin{equation}
 \rho^{\uparrow} = \sum_{\alpha,\beta}^{dim(\mathscr{H}_{\uparrow})}\sum_{m}\delta_{n(\alpha),n(\beta)}
 C_{\alpha,m}C_{\beta,m}^{*}\Ket{\alpha}\bra{\beta} \,,
 \label{reduced-density}
\end{equation}
where $n(\alpha)$ is the number of particles of the Fock state $\Ket{\alpha}$ 
in the subspace of the top ring,
$\mathscr{H}_{\uparrow}$.

This expression is valid for any system in which the bipartite splitting leads to 
two equal subsystems with the same dimension.  
For instance, consider $N$ bosons in a Josephson junction with two sites. 
The many-body state in the superfluid regime (when the tunneling is larger than the 
interaction) is given by:
\begin{equation}
 \Ket{\Psi_{\rm sf}}=\frac{1}{2^{N/2}}\sum_{n=0}^{N} \, \sqrt{\binom{N}{n}}\Ket{n,N-n} \,.
 \label{binomial}
\end{equation}
The bipartite splitting corresponds to each site. Therefore, the eigenstates of 
the subspace $\mathscr{H_{\uparrow}}$ are the Fock states of the top site
like $\Ket{n}$. 
The reduced density matrix $\rho^{\uparrow}$ is diagonal, whose eigenvalues (and the Schmidt spectrum) are given by 
the binomial distribution $\lambda_{k}=2^{-N}\binom{N}{n}$. This leads to the following analytical expressions 
for the Schmidt gap $\Delta \lambda^{\uparrow}=2^{-N}\left[\binom{N}{N/2}-\binom{N}{N/2-1}\right]$ (for $N$ even), 
and  the entanglement entropy $S_{\uparrow}=N\ln(2)-2^{-N}\sum_{k=0}^{N}\binom{N}{k}\ln \binom{N}{k}$.

\end{widetext}

\bibliography{refs.bib}
\end{document}